\documentclass[5p]{elsarticle}
\biboptions{sort&compress}

\usepackage{amssymb}
\usepackage{amsmath}
\usepackage{mathtools}
\usepackage{amsfonts}
\usepackage{bbm}
\usepackage{bbold}
\usepackage{algpseudocode}
\usepackage{algorithm}
\usepackage{braket}
\usepackage{caption}
\usepackage{subcaption}
\usepackage[utf8]{inputenc}
\usepackage[T1]{fontenc}
\usepackage{hyperref}
\hypersetup{
  colorlinks   = true, 
  urlcolor     = blue, 
  linkcolor    = blue, 
  citecolor   = blue 
}

\usepackage{tikz}

\usepackage[european]{circuitikz}
\usetikzlibrary{angles, quotes}
\usetikzlibrary{shapes.geometric}
\usetikzlibrary{shapes.misc, positioning}
\usetikzlibrary{matrix}
\usetikzlibrary{decorations.text}
\usetikzlibrary{backgrounds}
\usetikzlibrary{decorations.pathreplacing}
\usetikzlibrary{shapes.symbols}

\def\e{\ensuremath{\mathrm{e}}}
\def\i{\ensuremath{\mathrm{i}}}
\def\d{\ensuremath{\mathrm{d}}}
\DeclareMathOperator{\Tr}{Tr}

\newcommand{\mar}[1]{\textcolor{black}{#1}}
\newcommand{\new}[1]{\textcolor{black}{#1}}

\newcommand{\kket}[1]{|#1\rangle\rangle}

\journal{Computer Physics Communications}

\begin{document}

\begin{frontmatter}

\title{Numerical implementation of the partial secular approximation and unified master equation in structured open quantum systems}

\author[HY,Aalto]{Antti Vaaranta\corref{cor}}
\ead{antti.vaaranta@helsinki.fi}
\cortext[cor]{Corresponding author}
  
\author[HY]{Marco Cattaneo}
\ead{marco.cattaneo@helsinki.fi}

\affiliation[HY]{organization={QTF Centre of Excellence, Department of Physics, University of Helsinki},
  addressline={P.O. Box 43},
  city={FI-00014 Helsinki},
  country={Finland}}
\affiliation[Aalto]{organization={QTF Centre of Excellence, Department of Applied Physics, Aalto University},
  addressline={P.O. Box 15100},
  city={FI-00076 Aalto},
  country={Finland}}

\begin{abstract}
\mar{The Markovian dynamics of} open quantum systems is \mar{typically} described \mar{through} Lindblad equations, which are derived from the Redfield equation via the \textit{full} secular approximation. \mar{The latter neglects} the rotating terms in the master equation corresponding to \mar{pairs of} jump operators with different Bohr frequencies. However, for many physical systems this approximation breaks \mar{down}, and thus \mar{a} more accurate treatment of the \mar{slowly rotating terms} is \mar{required}. Indeed, more precise physical results can be obtained by performing the \textit{partial} secular approximation, which takes into account the relevant time scale \mar{associated with} each \mar{pair of} jump operators and compares \mar{it} with the time scale arising from the system-environment coupling. \mar{In this work, we introduce a general code} for performing the partial secular approximation \mar{in the Redfield equation for structured open quantum systems. The code can be applied to a generic Hamiltonian of any multipartite system} coupled to bosonic baths. \mar{Moreover, it can also reproduce the \textit{unified master equation}, which captures the same physical behavior as the Redfield equation under the partial secular approximation, but is mathematically guaranteed to generate a completely positive dynamical map. \mar{Finally, the code can compute both the local and global version of the master equation for the same physical problem.} We illustrate the code by studying the steady-state heat flow in a structured open quantum system composed of two superconducting qubits, each coupled to a bosonic mode, which in turn interacts with a thermal bath.} The results in this work can be \mar{employed for the numerical} study of a wide range of complex open quantum systems.
\end{abstract}




\end{frontmatter}

\section{Introduction} \label{intro}

The study of open quantum systems is critical for understanding the behaviour of the different emerging quantum technologies. Quantum processors and algorithms \mar{for} quantum computation \cite{Verstraete2009,Olivera2023,Vaaranta2022}, quantum heat engines and other applications \mar{in} quantum thermodynamics \cite{Pekola2015,Pekola2021,Alicki1979}, and \mar{accurate} sensors  \mar{in} quantum metrology \cite{Alipour2014,Montenegro2025,Jiao2025} all rely on \mar{the} careful characterization and manipulation of the quantum system of interest and its environment, which is at the heart of the theory of open quantum systems \cite{breuer2002theory}.

Master equations have a central role in the study of open quantum systems \cite{breuer2002theory}. They play the part of the Liouville-von Neumann equation \mar{for} closed quantum systems, evolving the initial quantum state, represented as a density matrix, in time under the influence of an environment. 
\mar{A physical time evolution must maintain} the positivity and unit trace of the density matrix. In the language \mar{of Markovian} open quantum systems, we say that the generator of the quantum dynamical semigroup
is required to be completely positive and trace preserving (CPTP). \mar{It can be formally shown that, under certain assumptions, a CPTP Markovian evolution is guaranteed if the coupling between the system and the environment is weak \cite{Davies1974}. In more detail, the property of}  complete positivity  \mar{guarantees} that the density matrix describing the \mar{open} quantum system of interest, \mar{which is} a subsystem of the total system where the environment is included, is \mar{positive at all times}. Trace preservation \mar{means that} the trace of the density matrix \mar{is always unity}. An equation with these properties, guaranteeing a CPTP \mar{Markovian} time evolution, is called a \textit{GKLS} master equation (after Gorini, Kossakowski, Sudarshan \cite{Gorini1976} and Lindblad \cite{Lindblad1976}), or just a \textit{Lindblad} equation.

In the derivation of the Lindblad equation several simplifying assumptions are made, restricting \mar{its} applicability.  \mar{The last step} in the microscopic derivation of the Lindblad equation \mar{is the so-called \textit{secular approximation}} \cite{Lidar2019,Manzano2020,Davies1974}.
\mar{More specifically, in the standard derivation of a microscopic master equation the \textit{full} secular approximation is performed. The latter} relies on the assumption that the \mar{relaxation time scale of the open system} $\tau_\text{\mar{R}}$ \mar{is much longer than the inverse of} the \mar{separation of any pair of different} Bohr frequencies\footnote{\mar{The Bohr frequencies are defined as the differences between two energy levels in the system Hamiltonian.}} $\tau_{\mar{\omega,\omega'}} = |\omega - \omega^\prime|^{-1}$, i.e., 
\begin{equation}
\label{eq:fullSec}
\tau_{\omega,\omega'} \ll \tau_\text{R}  \text{ for all } \omega\neq\omega'.
\end{equation}

 Although the \mar{application of} the full secular approximation \mar{guarantees} a completely positive and trace preserving dynamical map, the requirement \mar{in Eq.~\eqref{eq:fullSec}} is a strong assumption. \mar{In structured open quantum} systems it \mar{typically} does not hold, as  these can have arbitrarily small separation between the Bohr frequencies, leading to large $\tau_{\omega,\omega'}$. Then, the incorrect application of the full secular approximation \mar{neglects} dissipative terms in the master equation \mar{that} would give rise to  important \mar{physical} effects \cite{Wichterich2007,Jeske2015,Gonzalez2017}. 

To cure \mar{this issue}, the \textit{partial} secular approximation, \mar{that is, removing only the fast-rotating terms,} can be \mar{performed} instead. This \mar{was originally proposed by Redfield himself \cite{Redfield1965}, and it has been recently studied in a number of works \cite{Tscherbul2015,Jeske2015,Farina2019,Cattaneo2019,Dodin2024,Tscherbul2025}.}

\mar{A well-known drawback of the partial secular approximation is the fact that the resulting master equation is not guaranteed to be completely positive. While for most weak-coupling scenarios this may not be an issue for the sake of numerical simulations \cite{McCauley2020,Hartmann2020}, and partial secular equations are guaranteed to be completely positive for some specific models \cite{Farina2019,Tscherbul2025}, one may prefer to work with a master equation that is always mathematically well-defined. Then, different proposals for a master equation that is both in GKLS form and consistent with the predictions of the partial secular equation have been recently introduced, see for instance Refs.~\cite{Davidovic2020,Mozgunov2020a,Nathan2020,Trushechkin2021,Becker2021,DAbbruzzo2023,Scandi2025}.}
\mar{In this work, we take into account the proposal that is, perhaps, the most similar in spirit to the partial secular approximation, namely the \textit{unified master equation} introduced in Ref.~\cite{Trushechkin2021} (see also the related Refs.~\cite{Ptaszynski2019a,McCauley2020,Potts2021,Benhayounekhadraoui2025}). The unified master equation is based on creating different clusters of Bohr frequencies with similar magnitude, which are then treated as a single independent Bohr frequency in the master equation, so as to guarantee the GKLS form.}

In this \mar{paper} we provide \mar{a code for the} computational implementation of the partial secular approximation for a general \mar{structured open quantum system} coupled to an arbitrary number of bosonic baths. We \mar{describe the condition for the validity} of the partial secular approximation \mar{through the existence of a sufficiently long intermediate time scale between $\tau_{\omega,\omega'}$ and $\tau_\text{R}$}, as was formulated in Ref.~\cite{Cattaneo2019} \mar{(see also the general derivation in Ref.~\cite{Lidar2019})}. \mar{We put forward a simple algorithm that verifies this condition} numerically on a term-by-term basis. In addition, we also provide a numerical implementation of the \mar{the unified master equation} and discuss the \mar{feasibility} of this approach for general systems. \mar{Furthermore, the code can reproduce both local and global master equations \cite{Gonzalez2017,Cattaneo2019,Scali2021}, in partial secular, full secular, or unified regime}\new{, and also quantify the negativity of the obtained solutions.} Finally, \mar{our code can also exploit the} symmetries of the Liouvillian superoperator as a generator of the open quantum system dynamics \cite{Albert2014,Sa2023} \mar{to} block diagonalize \mar{its} matrix representation. This simplifies its structure, allowing for a more efficient computation of \mar{the} physical properties of the system. \mar{For instance, the Lindblad equation under the full secular approximation is symmetric with respect to the superoperator associated with the system Hamiltonian \cite{Lostaglio2017}. Moreover, also the partial secular approximation induces a symmetry of the Liouvillian under very general conditions, which can then be exploited for dimensionality reduction leading to an easier solution of the open system dynamics  \cite{Cattaneo2020}.}


\mar{We test our code} by computing the steady-state-heat flow through a \mar{system made of two superconducting qubits and two resonators coupled in a symmetric chain}, which is interacting with two thermal baths at its edges. In \mar{such a structured open quantum systems,} the separation \mar{between different} Bohr frequencies  \mar{is usually small enough that the full secular approximation is not justified anymore. Then,}
we study the difference between the \mar{so-called} local and global master equations \cite{Gonzalez2017,Cattaneo2019} with full and partial secular approximations. The \mar{numerical simulations identify} the \mar{regimes of parameters} where the local master equation \mar{and/or} full secular approximation fail. \mar{Moreover, we also present} a comparison between the term-by-term implementation of the partial secular approximation and the unified master equation, showing an agreement between the two approaches. The source code and other examples, \mar{with all the code documentation}, can be found in the public GitHub repository \cite{Vaaranta2025_GitHub}.

This manuscript is organized as follows. In Section \ref{sec:theor-backgr} we \mar{introduce} the necessary theoretical background to the open quantum systems formalism, starting from the Redfield equation to which the full secular approximation is usually applied. There, we also shortly discuss \mar{the} symmetries of the Liouvillian superoperator, which can be used to block diagonalize the Liouvillian matrix. In Section \ref{sec:algor-perf-part} we \mar{describe our algorithm} for performing the partial secular approximation \mar{and unified master equation,} together with a simple algorithm on how to obtain the block diagonal form of a general Liouvillian given that a symmetry exists. Section \ref{sec:examples} is dedicated to \mar{the example}, where we \mar{employ our code} to compute the steady-state heat flow through a qubit-resonator chain system. Finally, in Section \ref{sec:conclusions} we draw some concluding remarks.

\section{Theoretical background} \label{sec:theor-backgr}

\mar{In this section we introduce the theoretical tools for our analysis of different master equations.}
Throughout the manuscript we work with natural units, $\hbar = k_\text{B} = 1$.

We start by briefly reviewing the microscopic derivation of the master equation approach to open quantum systems in Section \ref{sec:redf-mast-equat}. There we focus on the Redfield equation and especially to the secular approximation that is usually applied to obtain a master equation of the Lindblad form. In Section \ref{sec:part-secul-appr} the partial secular approximation is introduced in detail and in Section \ref{sec:local-global-master} we discuss the global and local approaches to deriving the master equation and also mention some problems arising from naive application of the full secular approximation, which can be cured via the partial secular approximation. Then, \mar{the} unified master equation is discussed in Section \ref{sec:unif-mast-equat}

The final Section \ref{sec:symm-cons-quant} is devoted to introducing the theoretical background for exploiting the symmetries of the Liouvillian superoperator in order to \mar{study some specific physical quantities} without \mar{the need to solve} the full master equation, which can in general be either tricky or time consuming, especially for large system sizes.

\subsection{The Redfield \mar{and Lindblad} master equations} \label{sec:redf-mast-equat}

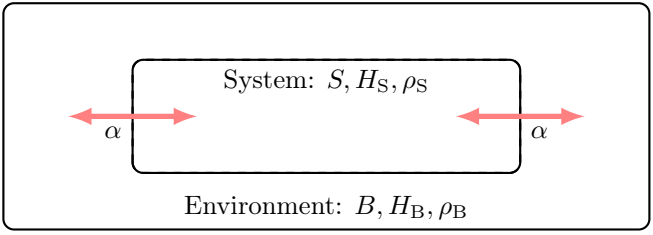
\begin{figure}
  \centering
  \begin{tikzpicture}[thick]
    \def\L{8.5}
    \def\H{3}

    \draw[rounded corners] (0, 0) rectangle (\L, \H);
    \node[anchor=south] at (\L/2,0) {Environment: $B, H_\text{B}, \rho_\text{B}$};

    \draw[rounded corners] (0.2*\L, 0.25*\H) rectangle (0.8*\L, 0.75*\H);    
    \draw[rounded corners, dashed] (0.2*\L, 0.25*\H) rectangle (0.8*\L, 0.75*\H);    
    \node[anchor=north] at (\L/2,0.75*\H) {System: $S, H_\text{S}, \rho_\text{S}$};
    
    \draw[<->, >=latex, red!50!white, line width=2pt] (0.1*\L, 0.5*\H) to (0.3*\L, 0.5*\H);
    \draw[<->, >=latex, red!50!white, line width=2pt] (0.7*\L, 0.5*\H) to (0.9*\L, 0.5*\H);
    \node[anchor=north] at (0.17*\L, 0.5*\H) {$\alpha$};
    \node[anchor=north] at (0.83*\L, 0.5*\H) {$\alpha$};
    
    \node[anchor=south] at (0.5*\L, \H) {Total system: $T, H_T = H_\text{S} + H_\text{B} + \alpha H_\text{I}, \rho_T=\rho_\text{S}\otimes\rho_\text{B}$};
  \end{tikzpicture}
  \caption{The system with Hamiltonian $H_\text{S}$ interacts with an environment with Hamiltonian $H_\text{B}$, making the system open. The interaction strength is given by the dimensionless coupling coefficient $\alpha$, assumed to be weak such that $\alpha\ll1$, and the interaction is mediated by the interaction Hamiltonian $H_\text{I}$. Figure adapted from \cite{breuer2002theory}.}
  \label{fig:OQS-schematic}
\end{figure}

The \mar{Markovian dynamics} of open quantum systems is governed by Lindblad or Gorini-Kossakowski-Sudarshan-Lindblad (GKLS) master equations, which describe the effect of an environment coupled weakly to some quantum system \cite{Gorini1976,Lindblad1976}. The total system-environment Hamiltonian can be written as
\begin{equation}
  \label{eq:Total-Ham}
  H_\text{T} = H_\text{S} + H_\text{B} + \alpha H_\text{I}\,,
\end{equation}
where $\alpha$ is a dimensionless coupling constant, see Fig. \ref{fig:OQS-schematic}. The subscripts $S$ and $B$ refer to the system and the environment (bath) respectively. The interaction part of the total Hamiltonian (\ref{eq:Total-Ham}) can be decomposed as
\begin{equation}
  \label{eq:H_I-def}
  H_\text{I} = \sum_\beta A_\beta \otimes B_\beta\,,
\end{equation}
where $A_\beta$ acts on the system's Hilbert space and $B_\beta$ to the Hilbert space of the environment. The index $\beta$ runs over all terms in the interaction Hamiltonian.

Here we focus on the microscopic derivation of the Lindblad master equation in the weak coupling limit between the system of interest and the environment. It is based on approximations grounded on the system's physical characteristics and, as a well known approach, it has been discussed in detail in many textbooks and scientific articles, for example in \cite{breuer2002theory,Rivas2010,Rivas2012,Lidar2019,Manzano2020,Davies1974}. \mar{We refer the readers to these references for a more thourough discussion. Here,} we only \mar{go through} the key elements for this procedure, which are \mar{listed below in the order of appearance in the usual derivation.}
\begin{enumerate}
\item \textbf{Weak coupling of the system to the environment:} $\alpha \ll 1$, where $\alpha$ is a dimensionless coupling coefficient multiplying the interaction Hamiltonian $\alpha H_\text{I} = \alpha\sum_\beta A_\beta\otimes B_\beta$, where $A_\beta$ is an operator acting on the Hilbert space of the system and $B_\beta$ is an operator acting on the Hilbert space of the environment.
\item \textbf{Born approximation:} \mar{The correlations between system and environment generated by the interaction Hamiltonian can be neglected in the derivation of the master equation, where we can replace the state of the total system at time $t$ with $\rho_\text{T}(t) = \rho_\text{S}(t)\otimes\rho_\text{B}(0)$. The initial state of the environment is an equilibrium state, such that $[H_\text{B}, \rho_\text{B}] = 0$.}
\item \textbf{Markovian time evolution:} The system's history can be neglected by assuming that the time scale of the correlations within the environment is much smaller that the time scale of the system's \mar{relaxation in interaction picture}, i.e. $\tau_\text{B} \ll \tau_\text{\mar{R}}$, 
\item \textbf{\mar{Full} secular approximation:} \mar{We neglect} the rotating terms \mar{in the dissipator of} the master equation in the interaction picture proportional to $\e^{-\i(\omega - \omega^\prime)t}$  with different Bohr frequencies, $\omega\neq\omega^\prime$.
\end{enumerate}

After performing \mar{these} four steps, one arrives at the standard form of the Lindblad equation, where the usual unitary dynamics \mar{described by} the Liouville-von Neumann equation \mar{is} complemented by a non-unitary contribution, \mar{usually referred to as the \textit{dissipator}}:
\begin{equation}
  \label{eq:Lindblad-equation}
  \begin{aligned}
    \dot{\rho}_\text{S} = &-\i[H_\text{S}+H_\text{LS}, \rho_\text{S}] \\
                          &+\sum_{\omega,\beta}\bigg(L_\beta(\omega)\rho_\text{S}L_\beta^\dagger(\omega)- \frac{1}{2}\left\{L_\beta^\dagger(\omega)L_\beta(\omega), \rho_\text{S}\right\}\bigg)\,.                            
  \end{aligned}
\end{equation}
Here $H_\text{LS}$ is the \textit{Lamb shift term}, \mar{which is a unitary contribution to the dynamics generated by the interaction with the environment}.
 $L_\beta(\omega)$ 
are the \textit{Lindblad operators}. They are modified forms of jump operators (which we will encounter soon), related to the Bohr frequencies.
 
Having performed the full secular approximation
, we are \mar{guaranteed} that the \mar{dynamics is described by a}  quantum dynamical semigroup \cite{Rivas2012}, preserving the trace and positivity of the density matrix, while simultaneously capturing the additional effects induced by the environment, such as dissipation and decoherence, on the quantum system of interest.

Now we take a step back from the above equation because, as said, obtaining Eq.~(\ref{eq:Lindblad-equation}) requires the  \textit{full} secular approximation on the different Bohr frequencies\mar{, which is not always justified. Indeed,} the difference \mar{between two Bohr frequencies} $|\omega-\omega^\prime|$ might be tiny but still nonzero. Therefore, we \mar{now go back} to the master equation after step 3 in the above list, called the Redfield equation, and perform step 4 in a more careful and considerate way.

The Redfield equation 
can be written in the interaction picture as \cite{Lidar2019,breuer2002theory}
\begin{equation}
  \label{eq:Bloch-Redfield-ME}
  \dot{\tilde{\rho}}_\text{S} = -\alpha^2\sum_{\substack{\beta, \beta^\prime \\ \omega, \omega^\prime}} \e^{-\i(\omega - \omega^\prime)t}\Gamma_{\beta \beta^\prime}(\omega)\left[A^\dagger_\beta(\omega^\prime), A_{\beta^\prime}(\omega)\tilde{\rho}_\text{S}\right] + \text{H.c.}
\end{equation}
with $\tilde{\rho}_\text{S}$ denoting the system density matrix in the interaction picture.

 $A_\beta(\omega)$ are called jump operators with \mar{Bohr} frequency $\omega$, defined via the relation
\begin{equation}
  \label{eq:jump-op-def}
  A_\beta(\omega)= \sum_{\epsilon_j - \epsilon_i = \omega}\braket{\epsilon_i|A_\beta|\epsilon_j}\ket{\epsilon_i}\bra{\epsilon_j}\,.
\end{equation}
Here $\ket{\epsilon_i}$ is the eigenstate of the system Hamiltonian $H_\text{S}$ with eigenenergy $\epsilon_i$. 
The jump operator with frequency $\omega$ is therefore related to the jumps between system's energy levels with Bohr frequency $\omega = \epsilon_j - \epsilon_i$.

In Eq.~(\ref{eq:Bloch-Redfield-ME}), the factor $\Gamma_{\beta\beta^\prime}(\omega)$ is the one-sided Fourier transform of the bath correlation function,
\begin{equation}
   \label{eq:one-sided-Fourier-of-bath-corr-func}
  \Gamma_{\beta \beta^\prime}(\omega) \equiv \int_0^\infty \d s \big\langle\tilde{B}_\beta^\dagger(s)\tilde{B}_{\beta^\prime}(0)\big\rangle \e^{\i\omega s} \,,
\end{equation}
where $\tilde{B}_\beta(t)$ is the (interaction-picture) bath operator that couples to the system operator $A_\beta$ in Eq.~(\ref{eq:H_I-def}).

Next, it is common to define \mar{the} quantities
\begin{subequations}
  \begin{align}
    \label{eq:gamma(omega,omega-prime)}
    \gamma_{\beta\beta^\prime}(\omega, \omega^\prime) &\equiv \Gamma_{\beta \beta^\prime}(\omega) + \Gamma_{\beta^\prime \beta}^*(\omega^\prime), \\
    \label{eq:Pi(omega,omega-prime)}
    \pi_{\beta\beta^\prime}(\omega, \omega^\prime) &\equiv \frac{1}{2\i}\left(\Gamma_{\beta \beta^\prime}(\omega) - \Gamma_{\beta^\prime \beta}^*(\omega^\prime)\right),
  \end{align}
\end{subequations}
and then to write the interaction picture equation (\ref{eq:Bloch-Redfield-ME}) in the Schrödinger picture \mar{as}
\begin{equation}
  \label{eq:Redfield-ME-Schröd-pic}
  \frac{\d}{\d t}\rho_\text{S}(t) = -\i\left[H_\text{S} + H_\text{LS}, \rho_\text{S}(t)\right] + \alpha^2\mathcal{D}[\rho_\text{S}(t)]\,,
\end{equation}
where the dissipator is given by
\begin{equation}
  \label{eq:dissipator-general}
  \begin{aligned}
    \mathcal{D}[\rho_\text{S}(t)] = \sum_{\beta, \beta^\prime}\sum_{\omega, \omega^\prime}&\gamma_{\beta\beta^\prime}(\omega, \omega^\prime)\bigg(A_{\beta^\prime}(\omega)\rho_\text{S}(t)A^\dagger_\beta(\omega^\prime) \\
                                                                                          &- \frac{1}{2}\left\{A^\dagger_{\beta}(\omega^\prime)A_{\beta^\prime}(\omega), \rho_\text{S}(t)\right\}\bigg)\,.
  \end{aligned}
\end{equation}
\mar{The Lamb-shift term of the Redfield equation} is defined as
\begin{equation}
  \label{eq:Lamb-shift-Hamiltonian}
  H_\text{LS} = \sum_{\beta, \beta^\prime}\sum_{\omega, \omega^\prime} \pi_{\beta\beta^\prime}(\omega, \omega^\prime)A^\dagger_{\beta}(\omega^\prime)A_{\beta^\prime}(\omega)\,.
\end{equation}

The full secular approximation discards \mar{all terms with $\omega\neq\omega'$ in the dissipator and Lamb-shift term of the Redfield equation, yielding Eq.~\eqref{eq:Lindblad-equation}}. 

\subsection{The partial secular approximation} \label{sec:part-secul-appr}

The justification for the secular approximation, whether full or partial, is the common rotating-wave-approximation -type argument. The idea is that the contribution of the fast rotating terms averages out to zero over sufficiently long time intervals as the underlying differential equation is integrated over time. This can be rigorously justified by defining a rescaled time $\tau = \alpha^2t$ and $\sigma = \alpha^2 s$, thus obtaining [via integrating Eq.~(\ref{eq:Bloch-Redfield-ME})]
\begin{multline}
  \label{eq:time-rescaled-me}
  \tilde{\rho}_\text{S}(\tau/\alpha^2) = \rho_\text{S}(0) - \sum_{\substack{\beta, \beta^\prime \\ \omega, \omega^\prime}} \int_0^{\tau}\d\sigma \e^{-\i(\omega - \omega^\prime)\sigma/\alpha^2}\Gamma_{\beta \beta^\prime}(\omega) \\
  \times\left[A^\dagger_\beta(\omega^\prime), A_{\beta^\prime}(\omega)\tilde{\rho}_\text{S}(\sigma/\alpha^2)\right] + \text{H.c.}\,.
\end{multline}
In the limit of $\alpha\to0$, while keeping $\tau$ and $\sigma$ constant, one can apply the Riemann-Lebesgue lemma (see chapter 6.2 of Ref.~\cite{Rivas2012} for a detailed discussion and \cite{Davies1974} for the original derivation) and fully justify the full secular approximation.

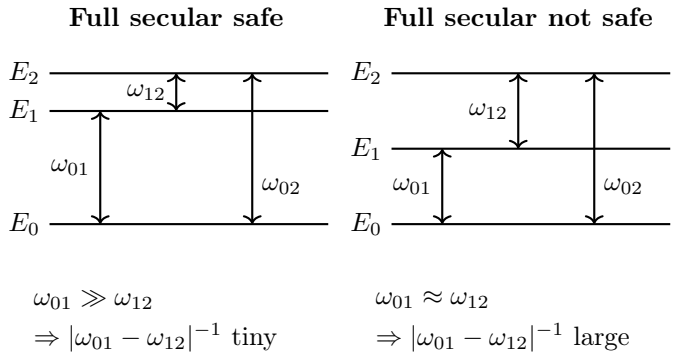
\begin{figure}
  \centering
  \begin{tikzpicture}

    \begin{scope}
      \node[anchor=south] at (2,2.5) {\textbf{Full secular safe}};
      \node at (0,0) (E0) {$E_0$};
      \node at (0,1.5) (E1) {$E_1$};
      \node at (0,2) (E2) {$E_2$};
      \draw[thick] (E0) -- (4,0);
      \draw[thick] (E1) -- (4,1.5);
      \draw[thick] (E2) -- (4,2);

      \draw[thick, <->] (1,0) -- (1,1.5) node[midway, left] {$\omega_{01}$};
      \draw[thick, <->] (2,1.5) -- (2,2) node[midway, left] {$\omega_{12}$};
      \draw[thick, <->] (3,0) -- (3,2) node[near start, right] {$\omega_{02}$};

      \node[anchor=west] at (0,-1) {$\omega_{01} \gg \omega_{12}$};
      \node[anchor=west] at (0,-1.5) {$\Rightarrow |\omega_{01} - \omega_{12}|^{-1}$ tiny};
    \end{scope}
    
    \begin{scope}[xshift=4.5cm]
      \node[anchor=south] at (2,2.5) {\textbf{Full secular not safe}};
      \node at (0,0) (E0b) {$E_0$};
      \node at (0,1) (E1b) {$E_1$};
      \node at (0,2) (E2b) {$E_2$};
      \draw[thick] (E0b) -- (4,0);
      \draw[thick] (E1b) -- (4,1);
      \draw[thick] (E2b) -- (4,2);

      \draw[thick, <->] (1,0) -- (1,1) node[midway, left] {$\omega_{01}$};
      \draw[thick, <->] (2,1) -- (2,2) node[midway, left] {$\omega_{12}$};
      \draw[thick, <->] (3,0) -- (3,2) node[near start, right] {$\omega_{02}$};

      \node[anchor=west] at (0,-1) {$\omega_{01} \approx \omega_{12}$};
      \node[anchor=west] at (0,-1.5) {$\Rightarrow |\omega_{01} - \omega_{12}|^{-1}$ large};
    \end{scope}
  \end{tikzpicture}
  \caption{\mar{Left: structure of energy levels of the system Hamiltonian for which}  the full secular approximation can be safely applied. \mar{The different Bohr frequencies are always well-separated, in such a way that their inverse is much smaller than the relaxation time scale. Right: in the structure of the energy levels}   two jump frequencies are very close to each other. In this case, \mar{their difference can be small}, which means that the full secular approximation cannot be directly applied.}
  \label{fig:secular-approx-safe-or-not}
\end{figure}

Of course, in actual physical systems taking the mathematical limit of vanishing coupling $\alpha\to 0$ is not justified, as we are interested in weak but non-zero interactions. Therefore, in order to justify the partial secular approximation, we need to compare the relevant time scales. In Eq.~(\ref{eq:time-rescaled-me}) the exponential factor rotates at a rate $\alpha^{-2}|\omega-\omega^\prime|$, which should be fast compared to the characteristic decay rate of the integration kernel $|\Gamma_{\beta \beta^\prime}(\omega)|$, giving the dissipative dynamics of the system \cite{Trushechkin2021}. Therefore we obtain
\begin{equation}
  \alpha^{-2}|\omega-\omega^\prime| \gg |\Gamma_{\beta \beta^\prime}(\omega)|
\end{equation}
which leads directly to
\begin{equation}
  \label{eq:PSA-requirement}
  \tau_{\omega,\omega'} = |\omega-\omega^\prime|^{-1} \ll \frac{1}{\alpha^2|\Gamma_{\beta \beta^\prime}(\omega)|} = \tau_\text{R}\,,
\end{equation}
\mar{where we have introduced the relaxation time scale of the system in interaction picture, $\tau_\text{R}$.} For two Bohr frequencies $\omega \approx \omega^\prime$ the time scale $\tau_{\omega,\omega'}$ becomes large such that it is comparable to $\tau_\text{R}$. This breaks the requirement for the secular approximation. See Fig.~\ref{fig:secular-approx-safe-or-not} for \mar{a} pictorial explanation.

The obtained requirement in Eq.~(\ref{eq:PSA-requirement}) is a refinement of the \mar{argument} given in Ref.~\cite{Cattaneo2019}, where the partial secular approximation is \mar{said to be valid} if there exists a separation of time scales \mar{in such a way that}
\begin{equation}
  \label{eq:secular-approx-def}
  \exists\, t^* \,\text{such that}\, |\omega - \omega^\prime|^{-1} \ll t^* \ll \tau_\text{R} = \mathcal{O}(\alpha^{-2})\,.
\end{equation}
In words, the separation between the expected time scale of the system \mar{relaxation} $\tau_\text{R}$ and the time scale of the rotating terms $|\omega - \omega^\prime|^{-1}$ is \mar{large enough} that \mar{an} intermediate time scale $t^*$ between them can be \mar{identified} \cite{Lidar2019}. 
If this condition \mar{holds} for some pair of frequencies $\omega, \omega^\prime$, then the corresponding cross terms in the dissipator \mar{and Lamb-shift} of Eq.~(\ref{eq:Bloch-Redfield-ME}) or (\ref{eq:Redfield-ME-Schröd-pic}) can be \mar{neglected}. 

For the \mar{sake of} numerical implementations of Eq.~(\ref{eq:PSA-requirement}), we need to determine the exact meaning of the ``much less than'' symbol. This is done by introducing \mar{the quantity} $C_\text{PSA}$, which acts as a cutoff parameter for the approximation. Namely, the secular approximation is performed (i.e. the cross term discarded) if the following inequality holds:
\begin{equation}
  \label{eq:psa-def}
  C_\text{PSA}|\omega - \omega^\prime|^{-1} < \tau_\text{R} = \mathcal{O}(\alpha^{-2})\,.
\end{equation}
This definition also allows us to assume $|\Gamma_{\beta \beta^\prime}(\omega)| = \mathcal{O}(1)$ without loss of generality as its exact value can be incorporated into the cut off coefficient $C_\text{PSA}$.

\mar{For instance,} choosing $C_\text{PSA} = 10^4$ ensures that there is at least four orders of magnitude between the Bohr frequency time scale $\tau_{\omega,\omega'}$ and the system time scale $\tau_\text{R}$. Note that the definition in Eq.~(\ref{eq:psa-def}) encodes also the full secular approximation, as setting $C_\text{PSA} = 0$ gives always a true statement for any $\omega\neq\omega^\prime$, thus neglecting every single cross term, which corresponds exactly to the full secular approximation. Moreover, setting $C_\text{PSA} = \infty$ will never neglect any cross terms and therefore \mar{corresponds} to the Redfield equation (\ref{eq:Redfield-ME-Schröd-pic}).


\subsection{Local and global master equations and the secular approximation} \label{sec:local-global-master}

Master equations \mar{for multipartite open systems} can be divided into two classes: global and local, which differ in the \mar{construction of their} jump operators. The jump operators of both master equations are derived as outlined in the previous section, computing them according to Eq.~(\ref{eq:jump-op-def}) \cite{breuer2002theory}. The difference lies in the system Hamiltonian \mar{employed in the derivation}, from which the eigenstates $\ket{\epsilon_i}$ are computed. In the derivation of the local master equation the inner-system interactions\footnote{\mar{Here, we are referring to any interaction between the subsystems of a multipartite system. For instance, in a system made of many qubits, we are referring to the qubit-qubit interactions.}} of the Hamiltonian $H_\text{S}$  are neglected such that the eigenstates $\ket{\epsilon_i}$ are computed using the free Hamiltonian \cite{Cattaneo2019}, whereas for the global case the full system Hamiltonian is used. As a result, the dissipators \mar{and Lamb-shift term} of the resulting local master equation have jump operators which act locally on the part of the system that is directly connected to the bath. \mar{Therefore, the inner system interactions appear only in the unitary part of the master equation through the term $-\i[H_\text{S},\rho]$, while they play no role in the dissipators and Lamb-shift.}  
This \mar{approximated master equation is typically valid} in the limit of weak inner-system couplings \cite{Trushechkin2016,Gonzalez2017,Cattaneo2019}.

\mar{In a multipartite open quantum system, the full secular approximation applied to a global master equation typically fails when the coupling between the subsystems is small, as the level splitting of the full system Hamiltonian is also small (i.e., quasidegeneracies appear). In contrast, in these scenarios the full secular approximation is usually valid for  local master equations that neglect the inner system interactions in the derivation of the master equation, and therefore treats the energy levels as degenerate \cite{Wichterich2007,Gonzalez2017}. For instance,} the full secular approximation applied to the global master equation can destroy key dynamical features of the time evolution, leading to the local master equation being more accurate in describing the transient dynamics for weakly interacting open quantum systems with separate baths \cite{Scali2021}. However, \mar{if there are common baths acting collectively on multiple subsystems,} then even in the derivation of the local master equation one needs to carefully consider the secular approximation\mar{, as the full secular may fail} \cite{Cattaneo2019}.

\subsection{The unified master equation and frequency grouping} \label{sec:unif-mast-equat}

Although the partial secular approximation fixes some issues \mar{that arise from an incorrect application of the} full secular approximation, such as the neglection of key dynamical features like quantum beats \cite{Cattaneo2019}, the resulting master equation \mar{may have problems related to} positivity and thermodynamic consistency \cite{Soret2022} \mar{(see also Ref.~\cite{Hartmann2020} for a different perspective)}. \mar{Then, a well-defined GKLS master equation that has the same regime of validity as the partial secular approximation may be desired. One proposal going towards this direction is the unified master equation \cite{Trushechkin2021}. In the unified equation, the full secular approximation is applied on some clusters of similar Bohr frequencies \cite{Trushechkin2021,Potts2021}, which are set equal in order to avoid the removal of slowly-rotating terms. The GKLS structure is then guaranteed by the full secular approximation. The unified equation preserves the positivity of the dynamics while being valid in the same regime as the partial secular equation. Moreover, it predicts a thermal Gibbs state with respect to a slightly modified system Hamiltonian \cite{Trushechkin2021}.}



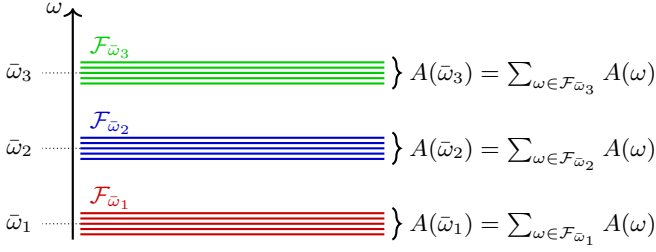
\begin{figure}
  \centering
  \begin{tikzpicture}
    \tikzstyle{every node}=[font=\small]
    \def \delta {0.07}
    \def \xlen {4}
    
    \foreach \i in {0,...,4}{
      \draw[thick, red!80!black] (0, \i*\delta) -- (\xlen, \i*\delta);
    }

    \foreach \i in {0,...,4}{
      \draw[thick, blue!80!black] (0, 1 + \i*\delta) -- (\xlen,  1 + \i*\delta);
    }

    \foreach \i in {0,...,4}{
      \draw[thick, green!80!black] (0, 2 + \i*\delta) -- (\xlen,  2 + \i*\delta);
    }

    \draw[densely dotted] (0, 2*\delta) -- (-0.5, 2*\delta) node[anchor=east] {$\bar{\omega}_1$};
    \draw[densely dotted] (0, 1 + 2*\delta) -- (-0.5, 1 + 2*\delta) node[anchor=east] {$\bar{\omega}_2$};
    \draw[densely dotted] (0, 2 + 2*\delta) -- (-0.5, 2 + 2*\delta) node[anchor=east] {$\bar{\omega}_3$};

    \draw[decorate, decoration={brace, amplitude=3pt, mirror, raise=3pt}, thick] (\xlen, -\delta) -- (\xlen, 5*\delta)
    node[anchor=west, yshift=-4pt] at (\xlen+0.2, 3*\delta) {$A(\bar{\omega}_1) = \sum_{\omega \in \mathcal{F}_{\bar{\omega}_1}}A(\omega)$};
    \draw[decorate, decoration={brace, amplitude=3pt, mirror, raise=3pt}, thick] (\xlen, 1-\delta) -- (\xlen, 1+5*\delta)
    node[anchor=west, yshift=-4pt] at (\xlen+0.2, 1+3*\delta) {$A(\bar{\omega}_2) = \sum_{\omega \in \mathcal{F}_{\bar{\omega}_2}}A(\omega)$};
    \draw[decorate, decoration={brace, amplitude=3pt, mirror, raise=3pt}, thick] (\xlen, 2-\delta) -- (\xlen, 2+5*\delta)
    node[anchor=west, yshift=-4pt] at (\xlen+0.2, 2+3*\delta) {$A(\bar{\omega}_3) = \sum_{\omega \in \mathcal{F}_{\bar{\omega}_3}}A(\omega)$};

    \draw[->, thick] (-0.1, -\delta) -- (-0.1, 3) node[anchor=east] {$\omega$};

    \draw[red!80!black] (0, 4*\delta) node[anchor=south west, yshift=-2pt] {$\mathcal{F}_{\bar{\omega}_1}$};
    \draw[blue!80!black] (0, 1+4*\delta) node[anchor=south west, yshift=-2pt] {$\mathcal{F}_{\bar{\omega}_2}$};
    \draw[green!80!black] (0, 2+4*\delta) node[anchor=south west, yshift=-2pt] {$\mathcal{F}_{\bar{\omega}_3}$};
    
  \end{tikzpicture}
  \caption{Pictorial representation of the Bohr frequency grouping. Each cluster $\mathcal{F}_{\bar{\omega}}$ is drawn with different color and they are clearly separated from each other. A cluster frequency $\bar{\omega}$ is selected to be the average Bohr frequency of each cluster. The jump operator of a cluster $A(\bar{\omega})$ is given by the sum of all jump operators $A(\omega)$ whose frequencies $\omega$ belong to that cluster.}
  \label{fig:freq-group-pic}
\end{figure}

The core idea of the frequency grouping is to divide the Bohr frequencies into well separated clusters such that the \mar{full} secular approximation is performed between elements from different clusters \mar{only} (see Fig.~\ref{fig:freq-group-pic}). \mar{All Bohr frequencies within the same cluster are assumed to have a similar magnitude, and are then set to the same value, which is} \new{for example} their average, in the final expression of the master equation.

A detailed and rigorous derivation of \mar{the unified master equation can be found} in \cite{Trushechkin2021}. Here, we \mar{just} outline the basic steps to \mar{group the Bohr frequencies in a proper way}. We start by decomposing the system Hamiltonian into two parts,
\begin{equation}
  \label{eq:H_S-decomp}
  H_\text{S} = H_\text{S}^{(0)} + \alpha^2\delta H_\text{S}\,.
\end{equation}
In $H_\text{S}^{(0)}$ the nearly degenerate Bohr frequencies of $H_\text{S}$ are exactly degenerate (giving the separation between different clusters) while the Bohr frequencies of $\alpha^2\delta H_\text{S}$ correspond to the quasidegeneracies. Then any Bohr frequency of the full Hamiltonian can be written as
\begin{equation}
  \omega = \bar{\omega} + \alpha^2\delta\omega\,,
\end{equation}
where $\bar{\omega}$ is a Bohr frequency of $H_\text{S}^{(0)}$ and $\delta\omega$ is a Bohr frequency of $\delta H_\text{S}$. The full set of Bohr frequencies is divided into clusters of Bohr frequencies, each having a distinctive cluster frequency $\bar{\omega}$ and other frequencies being around $\alpha^2$ apart from it, see Fig.~\ref{fig:freq-group-pic}.

Notice that indeed the cluster frequencies $\bar{\omega}$ of different clusters are well separated from each other while those from the same cluster are not:
\begin{equation}
  \omega - \omega^\prime = \bar{\omega} - \bar{\omega}^\prime + \alpha^2(\delta\omega - \delta\omega^\prime)\,.
\end{equation}
This means that the rotating exponential $\e^{-\i(\omega-\omega^\prime)t}$, in the interaction picture form of the Redfield equation (\ref{eq:Bloch-Redfield-ME}), is rapidly oscillating if and only if $\omega$ and $\omega^\prime$ belong to different clusters, i.e. for $\bar{\omega} \neq \bar{\omega}^\prime$. Otherwise (for $\bar{\omega} = \bar{\omega}^\prime$, i.e. $\omega$ and $\omega^\prime$ belonging to the same cluster) the time scale of the oscillations of the exponential factor is proportional to $\alpha^{-2}$, which is the same order as the system's time scale $\tau_\text{R}$, as defined in Eq.~(\ref{eq:secular-approx-def}).

This procedure therefore allows us to drop the cross terms with frequencies coming from different clusters $\bar{\omega} \neq \bar{\omega}^\prime$ while retaining the cross terms with frequencies in the same cluster $\bar{\omega} = \bar{\omega}^\prime$ due to rotation of the factor $\e^{-\i(\omega - \omega^\prime)t}$. In essence, we perform full secular approximation with respect to $H_\text{S}^{(0)}$, which is partial secular approximation with respect to $H_\text{S}$.

We can apply the above procedure at the level of the Redfield master equation (\ref{eq:Redfield-ME-Schröd-pic}). The spectrum of the Bohr frequencies of the system Hamiltonian $H_\text{S}$ are divided into clusters $\mathcal{F}_{\bar{\omega}}$ and for each a single cluster frequency $\bar{\omega}$ is picked (for example the average frequency within the cluster). Then the full secular approximation is performed with respect to the different cluster frequencies and we obtain the unified master equation:
\begin{equation}
  \label{eq:Unified-Redfield-ME-Schröd-pic}
  \frac{\d}{\d t}\rho_\text{S}(t) = -\i\left[H_\text{S} + H_\text{LS}, \rho_\text{S}(t)\right] + \alpha^2\bar{\mathcal{D}}[\rho_\text{S}(t)] \,,
\end{equation}
where the unified dissipator $\bar{\mathcal{D}}[\rho_\text{S}(t)]$ is given by
\begin{equation}
  \label{eq:dissipator-unified-me}
  \begin{aligned}
    \bar{\mathcal{D}}[\rho_\text{S}(t)] = \sum_{\beta, \beta^\prime}\sum_{\bar{\omega} \in \mathcal{F}^{(0)}}&\gamma_{\beta\beta^\prime}(\bar{\omega})\bigg(A_{\beta^\prime}(\bar{\omega})\rho_\text{S}(t)A^\dagger_\beta(\bar{\omega}) \\
    &- \frac{1}{2}\left\{A^\dagger_{\beta}(\bar{\omega})A_{\beta^\prime}(\bar{\omega}), \rho_\text{S}(t)\right\}\bigg)\,,
  \end{aligned}
\end{equation}
with
\begin{equation}
  A_\beta(\bar{\omega}) = \sum_{\omega\in\mathcal{F}_{\bar{\omega}}}A_\beta(\omega)\,.
\end{equation}
This shows that the jump operators in Eq.~(\ref{eq:dissipator-unified-me}) are sums of the jump operators belonging to the same cluster, see again Fig.~\ref{fig:freq-group-pic}. Now the factor $\gamma_{\beta\beta^\prime}(\bar{\omega})$ \mar{replaces} $\gamma_{\beta\beta^\prime}(\bar{\omega}, \bar{\omega})$ of Eq.~(\ref{eq:gamma(omega,omega-prime)}), which becomes the full Fourier transformation of the bath correlation functions:
\begin{equation}
  \label{eq:gamma-omega-bar}
  \gamma_{\beta\beta^\prime}(\bar{\omega}) = \int_{-\infty}^\infty \d \tau \big\langle\tilde{B}_\beta^\dagger(\tau)\tilde{B}_{\beta^\prime}(0)\big\rangle \e^{\i\bar{\omega} \tau}\,.
\end{equation}

\mar{To obtain the unified master equation we need to assume that the frequency} clustering can be performed in the first place. \mar{In other words, the structure of the energy levels must give rise to well-separated clusters of frequencies.} For small systems this  is usually \mar{the case}, but it comes as no surprise that \mar{in many-body systems we often cannot derive the unified equation, as they may display a dense range of Bohr frequencies. In these scenarios, creating well-separated clusters would necessarily mean that we are neglecting cross terms between jump operators with similar Bohr frequencies, violating the condition for the partial secular approximation. We discuss} these cases in more detail in Section \ref{sec:algor-perf-part}, where \mar{we introduce the code for the unified master equation.} 

\new{If the system frequencies can be grouped in well-defined clusters, then the regime of validity of the unified master equation is equivalent to that of the global master equation with partial secular approximation \cite{Trushechkin2021,Cattaneo2019}, with the additional benefit that the unified equation is always completely positive.}

\subsection{Symmetries \mar{in open quantum systems}} \label{sec:symm-cons-quant}

Solving the master equation, be it in the Lindblad, Redfield or some other form, is \mar{quite often} numerically  demanding due to the large size of the Hilbert space involved. 
To counteract the increase in computation time required for solving the full Lindblad dynamics, we can, in some cases, exploit the symmetries of the Liouvillian superoperator in order to obtain interesting physical results \mar{in} a much smaller subspace of the full Hilbert space. \mar{With} this goal in mind, we introduce the Liouville space (see, e.g., \cite{Gyamfi2020}) representation of Eq.~(\ref{eq:Redfield-ME-Schröd-pic}), given by
\begin{equation}
  \frac{\d}{\d t}\kket{\rho_\text{S}(t)} = \mathcal{L}\kket{\rho_\text{S}(t)} \Rightarrow \kket{\rho_\text{S}(t)} = \e^{\mathcal{L}t}\kket{\rho_\text{S}(0)}\,,
\end{equation}
where $\kket{\rho_\text{S}(t)}$ denotes the reshaped form of the density matrix, now being a state vector, and $\mathcal{L}$ is the similarly reshaped form of the Liouville superoperator, now represented as a matrix
\begin{equation}
  \label{eq:Liouvillian-sup-op-matrix}
  \mathcal{L} = -\i\left(H\otimes\mathbb{1} - \mathbb{1}\otimes H^T\right) + \alpha^2\mathcal{D}\,,
\end{equation}
where the Hamiltonian is $H = H_\text{S} + H_\text{LS}$ and the matrix form of the dissipator is given by
\begin{equation}
  \label{eq:dissipator-matrix}
  \begin{aligned}
    \mathcal{D} = \sum_{\beta,\beta^\prime}\sum_{\omega,\omega^\prime}&\gamma_{\beta\beta^\prime}(\omega, \omega^\prime)\bigg(A_{\beta^\prime}(\omega)\otimes A_\beta^*(\omega^\prime) \\
                                                                      &- \frac{1}{2}\left\{\!\!\left\{A_\beta^\dagger(\omega^\prime)A_{\beta^\prime}(\omega)\,,\mathbb{1}\right\}\!\!\right\}\bigg)\,,
  \end{aligned}
\end{equation}
where we used the superanticommutator defined as
\begin{equation}
  \label{eq:superanticomm}
  \left\{\!\!\left\{X, Y\right\}\!\!\right\} = X\otimes Y^T + Y\otimes X^T\,.
\end{equation}
It is the matrix form $\mathcal{L}$ of the Liouvillian that we wish to simplify by exploiting any symmetries present.


Following the discussion in \cite{Albert2014}, 
we say that an operator $J$ generates a continuous symmetry (one-parameter group) at the superoperator level, \mar{or weak symmetry, if}
\begin{equation}
  \label{eq:contiuos-symmetry}
  \e^{-\i\phi \mathcal{J}}\mathcal{L}\e^{\i\phi \mathcal{J}} = \mathcal{L}
\end{equation}
for $\phi\in\mathbb{R}$, or equivalently $[\mathcal{J}, \mathcal{L}] = 0$, where the superoperator form of $J$ is given by
\begin{equation}
  \label{eq:sup-op-J}
  \mathcal{J} = [J, \cdot]\; \mar{\rightarrow} \;J\otimes\mathbb{1} - \mathbb{1}\otimes J^T
\end{equation}
in the Liouville space formulation.

It should be noted that, in the case of \mar{the unitary dynamics in a} closed quantum system, \mar{the presence of a continuous symmetry implies the emergence of a conserved quantity of the dynamics.} However, in the case of open quantum systems having a symmetry at the superoperator level does not quarantee a symmetry at the operator level, which also implies that there does not necessarily exist a conserved quantity \cite{Albert2014}.

\mar{The symmerty property  $[\mathcal{J}, \mathcal{L}] = 0$ at the superoperator level implies that we can block diagonalize the Liouvillian matrix such that each block corresponds to a different eigenvalue of $\mathcal{J}$. In this way, the effective dimensionality of the linear system of equations we have to solve can be greatly reduced.}

\mar{Importantly, the Lindblad master equation in full secular approximation, Eq.~\eqref{eq:Lindblad-equation}, driven by the Liouvillian $\mathcal{L}$, always satisfies the symmetry
\begin{equation}
    [\mathcal{H}_S,\mathcal{L}]=0,
\end{equation}
where 
\begin{equation}
    \mathcal{H}_S = [H_S,\cdot],
\end{equation}
and $H_S$ is the full system Hamiltonian. Therefore, the Liouvillian can be block-diagonalized using the eigenstates of $H_S$ as a basis for the space of operators, i.e., assembled in any possible combination of kets and bras. This condition is a direct consequence of the full secular approximation \cite{Lostaglio2017}.}


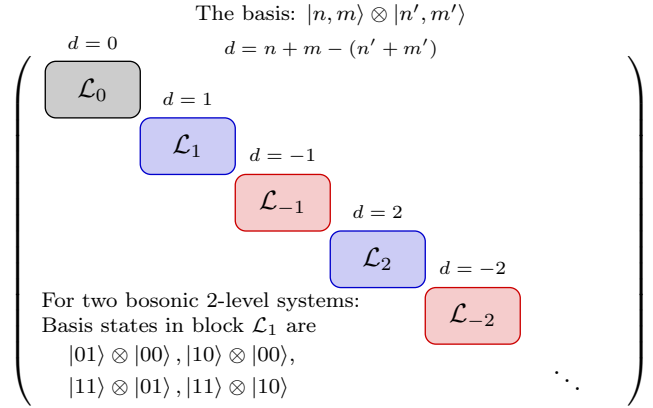
\begin{figure}
  \centering
  \begin{tikzpicture}

    \matrix[matrix of math nodes,
    left delimiter=(,
    right delimiter=),
    column sep=-0.5pt,
    row sep=-0.5pt,
    every node/.style={minimum width=1.25cm, minimum height=0.75cm},
    G/.style={draw=black, line width=0.5pt, fill=black!20, rounded corners},
    R/.style={draw=red!80!black, line width=0.5pt, fill=red!80!black!20, rounded corners},
    B/.style={draw=blue!80!black, line width=0.5pt, fill=blue!80!black!20, rounded corners}]
    (L){
      |[G]| $\mathcal{L}_{0}$ &  &  &  &  & \\
      & |[B]| $\mathcal{L}_{1}$ &  &  &  &  \\
      &  & |[R]| $\mathcal{L}_{-1}$ &  &  &  \\
      &  &  & |[B]| $\mathcal{L}_2$ &  &  \\
      &  &  &  & |[R]| $\mathcal{L}_{-2}$ &  \\
      &  &  &  &  &  $\ddots$ \\};

    \node[yshift=60pt] (a) at (L-3-3.north east) {\footnotesize The basis: $\ket{n,m}\otimes\ket{n',m'}$};
    \node[yshift=-5pt] at (a.south) {\scriptsize $d = n+m - (n'+m')$};
    
    \node[yshift=7pt] at (L-1-1.north) {\scriptsize$d=0$};
    \node[yshift=7pt] at (L-2-2.north) {\scriptsize$d=1$};
    \node[yshift=7pt] at (L-3-3.north) {\scriptsize$d=-1$};
    \node[yshift=7pt] at (L-4-4.north) {\scriptsize$d=2$};
    \node[yshift=7pt] at (L-5-5.north) {\scriptsize$d=-2$};

    \node[xshift=15pt, yshift=-5pt, above right, anchor=east] at (L-4-4.south west) {\footnotesize For two bosonic 2-level systems:};
    \node[below left, xshift=-37pt, yshift=14pt] (b) at (L-5-5.south west) {\footnotesize Basis states in block $\mathcal{L}_1$ are};
    \node[yshift=3pt, xshift=10pt, below right, align=left] at (b.south west) {\footnotesize $\ket{01}\otimes\ket{00}, \ket{10}\otimes\ket{00}$, \\ \footnotesize $\ket{11}\otimes\ket{01}, \ket{11}\otimes\ket{10}$};
    
  \end{tikzpicture}
  \caption{Block structure of the Liouvillian superoperator arising from the symmetry between the $\mathcal{L}$ and the total number of particles superoperator $\mathcal{N}$. The Liouvillian is block diagonalized such that each block can be labeled by an eigenvalue $d$ of $\mathcal{N}$. \mar{Here, $d$ is equal to the difference in the number of particles  between the left and right sides of the tensor product structure of the basis states  $\ket{n,m}\otimes\ket{n', m'}$ in Liouvillian space}. Figure modified from \cite{Vaaranta2022}.}
  \label{fig:L-block}
\end{figure}

\mar{The partial secular approximation breaks the symmetry with respect to $\mathcal{H}_S$. However, under very general assumptions it can be shown that a new symmetry arises \cite{Cattaneo2020}. Namely, if the system Hamiltonian can be written as a collection of many free quasi-particles, which can be bosonic, fermionic, or qubits, and $N$ denotes the total number of excitations in the system, then in many scenarios (see \cite{Cattaneo2020} for more details)
\begin{equation}
    [\mathcal{N},\mathcal{L}]=0\text{, with }\mathcal{N}=[N,\cdot]. 
\end{equation}}

As an example of this symmetry, let us consider a two-bosons system Hamiltonian which can be written as
\begin{equation}
  H=\omega_1 a_1^\dagger a_2 + \omega_2 a_2^\dagger a_2,
\end{equation}
where $a_1$ and $a_2$ are bosonic annihilation operators.
We construct the total-number-of-particles operator $N$ as
\begin{equation}
  N = a_1^\dagger a_1 + a_2^\dagger a_2\,.
\end{equation}

Let us then write its superoperator form as $\mathcal{N} = N\otimes\mathbb{1} - \mathbb{1}\otimes N$. \mar{If the system couples linearly to a generic Markovian thermal bath, we can derive a master equation in partial secular approximation, according to Eq.~\eqref{eq:Bloch-Redfield-ME}, whose Liouvillian superoperator commutes} with the total-number-of-particles superoperator, i.e. $[\mathcal{L},\mathcal{N}] = 0$, thus inducing a symmetry at the superoperator level \cite{Cattaneo2020}. \new{In this example, we have assumed that the frequencies of both bosonic modes $\omega_1$ and $\omega_2$ have the same order of magnitude and are much larger than the system-bath coupling. As a consequence, the partial secular approximation cancels the terms in the Liouvillian of the form $a_j \rho a_k$, for $j,k=1,2$. However, if $\omega_1\approx \omega_2$, it keeps the cross-terms such as $a_1 \rho a_2^\dagger$.}

\mar{If we represent the two-bosons state in the Fock space as $\ket{n,m}$, meaning that the first boson has $n$ excitations while the second $m$,} the Liouvillian can be block-diagona-lized in the basis $\ket{n,m}\otimes\ket{n',m'}$ such that the blocks are labeled by the eigenvalues $d$ of $\mathcal{N}$. These eigenvalues have a well defined meaning, as they correspond to the difference in the number of particles of the two kets in the Liouville space basis, $d=n+m-(n'+m')$. See Fig.~\ref{fig:L-block} for a pictorial representation of the block diagonalization.

\mar{Finally, it can be shown that the unified master equation is symmetric with respect to the superoperator associated with the Hamiltonian $H_S^{(0)}$ defined in Eq.~\eqref{eq:H_S-decomp} \cite{Trushechkin2021}.}


\section{Algorithms for performing the partial secular approximation \mar{and unified master equation}}\label{sec:algor-perf-part}
The master equations  \mar{for the open dynamics of even few-body quantum systems} are \mar{quite often} too difficult to solve analytically, except in some simple cases. Therefore, \mar{one usually has  to resort to numerical methods}. \mar{With} this in mind, in this section we introduce an algorithm for performing the secular approximation \mar{and unified master equation} for both the global and local cases.

The general outline for creating the master equation and solving for the dynamics numerically is:
\begin{enumerate}
\item \textbf{Determine the system Hamiltonian.} The \new{formula} of the Hamiltonian needs to be determined by the user.
\item \textbf{Determine the baths.} The baths are determined via their temperature, the \new{user-specified} spectral density, and the coupling to the system. The user also needs to  provide the system operators $A_\beta$, which \mar{are coupled} to the bath operators [see Eq.~(\ref{eq:H_I-def})]. The baths are assumed to be bosonic.
\item \textbf{Hamiltonian diagonalization.} This determines the eigenvectors and -energies, which are used to build the jump operators for the master equation derivation. This step is performed by the \mar{code}.
\item \textbf{The secular approximation.} This can be done in two ways:
  \begin{enumerate}
  \item Perform the partial secular approximation term-by-term by comparing the time scales of the Bohr frequency differences $|\omega-\omega^\prime|^{-1}$ to the estimated time scale of the dynamics $$\tau_\text{R} = \alpha^{-2}|\Gamma_{\beta \beta^\prime}(\omega)|^{-1},$$ implementing Eq.~(\ref{eq:psa-def}).
  \item Derive a unified master equation where all near degenerate Bohr-frequencies are collected into clusters with respect to which the full secular approximation is performed.
  \end{enumerate}
\item \textbf{Build the Liouvillian.} The Liouvillian $\mathcal{L}$ is constructed from the jump operator pairs which survive the secular approximation step, according to Eq.~(\ref{eq:Redfield-ME-Schröd-pic}).
\item \mar{\textbf{Solve for the dynamics.} Given the Liouvillian $\mathcal{L}$ and an initial state $\rho_S(0)$, compute the evolved state at time $t$ through
\begin{equation}
    \rho_S(t)=\exp(\mathcal{L}t)[\rho_S(0)]. 
\end{equation}
The Liouvillian is constructed as a matrix to be applied to the vectorized version of $\rho_S(0)$.}
\end{enumerate}

For the first two steps one needs to define the operators that appear on the Hamiltonian. These can be done by using the Python library QuTiP \cite{Johansson2012,Johansson2013}. Examples on how to define the Hamiltonian and the baths can be found in the \mar{Github repository} \cite{Vaaranta2025_GitHub}. \new{Additionally, the users can provide their own spectral density function for the bath. The function must have three arguments, $J(\omega, \chi, \omega_\text{c})$, being respectively the frequency of the jump, the coupling constant to the environment, and the cut-off frequency. As a default, the spectral density function is taken to be Ohmic with Drude cut off (see \ref{sec:coupl-boson-bath}).}

\mar{In what follows, we discuss the details of how the secular approximation and the frequency grouping in the unified master equation are performed in the code.} 

\subsection{An algorithm for the term-by-term partial secular approximation}
The pseudo-code for performing the partial secular approximation term-by-term is described  in Algorithm \ref{alg:do-secular-approx} below. \mar{It automatically establishes which cross terms are to be discarded according to the partial secular approximation,}  based solely \mar{on} the comparison \mar{between} the time scales $\tau_{\omega,\omega'} = |\omega - \omega^\prime|^{-1}$ and $\tau_\text{R} = \alpha^{-2}|\Gamma_{\beta \beta^\prime}(\omega)|^{-1}$. The \mar{system-bath} coupling $\alpha$ is \mar{an input variable of the algorithm}. The Bohr frequencies $\omega$ are instead calculated by the code while constructing the jump operators according to Eq.~(\ref{eq:jump-op-def}). Additionally, the user has to \mar{provide} the cutoff value $C_\text{PSA}$, which determines the confidence with which the partial secular approximation is applied: the larger $C_\text{PSA}$ is, the larger needs to be the difference between the two compared time scales in order for the term to be discarded. Therefore, as noted before in Section \ref{sec:part-secul-appr}, if $C_\text{PSA} = \infty$, none of the terms are discarded and we will obtain a Redfield equation. Similarly, if $C_\text{PSA} = 0$ all the cross terms with $\omega\neq\omega^\prime$ are neglected and we obtain the full secular approximation.

Algorithm \ref{alg:do-secular-approx} checks a single cross term, returning either true or false, \mar{which corresponds to whether the cross term under consideration} will \mar{respectively} be neglected or not.  
Essentially, Algorithm \ref{alg:do-secular-approx} goes through all the terms of the Redfield equation in Eq.~(\ref{eq:Redfield-ME-Schröd-pic}) and decides, based on the two frequencies $\omega$ and $\omega^\prime$, if each cross term should be kept or not. 

\mar{The algorithm also allows for the generation of a Liouvillian superoperator corresponding to a local master equation. To do so, the user has to pass the bare system Hamiltonian (i.e., without inner-system interactions) to the code. The jump operators and the dissipator are then computed with respect to this Hamiltonian, instead of the full system Hamiltonian, which appears only in the unitary part of the Liouvillian, as explained in Sec.~\ref{sec:local-global-master}.}

\begin{algorithm}[h]
  \caption{Do secular approximation}\label{alg:do-secular-approx}
  \begin{algorithmic}
    \Require $\omega_1, \omega_2 \in \mathbb{R}$ the jump operator frequencies; \newline$\tau_\text{R} > 0$ the coarse time scale of the system; \newline$C_\text{PSA} \geq 0$ the partial secular approximation cutoff
    \State $\Delta \gets |\omega_1 - \omega_2|$

    \If{$\Delta = 0$}
    \State $\tau_{\omega,\omega'} \gets \infty$
    \Else
    \State $\tau_{\omega,\omega'} \gets 1/\Delta$
    \EndIf

    \If{$\tau_{\omega,\omega'} C_\text{PSA} < \tau_\text{R}$}
    \Comment{Implements Eq.~(\ref{eq:psa-def})}
    \State dropTerm $\gets$ True
    \Else
    \State dropTerm $\gets$ False
    \EndIf

    \State \Return dropTerm
  \end{algorithmic}
\end{algorithm}

\subsection{An algorithm for the \mar{unified master equation}}

The frequency clustering \mar{of the unified master equation, introduced in Section~\ref{sec:unif-mast-equat},} is outlined  in Algorithm \ref{alg:freq-clustering}. This algorithm works best for a spectrum of Bohr frequencies  
which has well separated gaps. If the spectrum is flat with equal spacing between the Bohr frequencies, there are no definite clusters to be found and the algorithm leads into the Redfield equation where no jump operator pairs are discarded.

\mar{The algorithm requires a parameter  $w>0$ as input}, which describes the maximum value \mar{for the difference between two consecutive Bohr frequencies to be included in the same cluster}. Then, \mar{after sorting the Bohr frequencies in ascending order,} the algorithm compares the frequencies $\omega_i$ and $\omega_{i+1}$. If $\omega_{i+1} - \omega_i < w$, the two frequencies belong to the same cluster. Otherwise \mar{a new cluster is created and $\omega_{i+1}$ is assigned to this cluster}. \new{Once a cluster is completed, the frequency associated to it is  the average of all Bohr frequencies in the cluster.} A similar procedure was applied in Ref.~\cite{Benhayounekhadraoui2025}.

\mar{Note that, using this algorithm, the widths of the clusters may actually be larger than the separation between them. For instance,} let $C_1$ and $C_2$ be two neighbouring clusters with $\omega^\prime > \omega$ for all $\omega^\prime \in C_2$ and $\omega \in C_1$. Then let $\omega_{\min, 1} = \min\{C_1\}\,, \omega_{\max, 1} = \max\{C_1\}$ and $\omega_{\min,2} = \min\{C_2\}$. We may have $\omega_{\max,1}-\omega_{\min,1} > \omega_{\min,2}-\omega_{\max,1}$ if the Bohr frequency spectrum is not well suited for the clustering, see Fig.~\ref{fig:Bohr-spectra}. There the Bohr frequency clustering is done in the case of the qubit-resonator chain Hamiltonian introduced in Section \ref{sec:intr-qubit-reson}. On the left the coupling between the qubits and resonators is weak, which translates into well-defined clusters in the Bohr frequency spectrum. As the couplings \mar{increase} (right panel of Fig.~\ref{fig:Bohr-spectra}), the spectrum loses its structure, such that distinct clusters cannot be easily distinguished anymore. 

Algorithm \ref{alg:freq-clustering} returns the Bohr frequency clusters, which are then passed to the function that computes the Liouvillian superoperator. With the clustering, the function for the Liouvillian performs full secular approximation between all the frequencies that belong to different clusters, while the jump operators corresponding to frequencies belonging to the same cluster are summed together and the cluster frequency $\bar{\omega}$, being the average of all frequencies within a cluster, is used when computing the coefficient $\gamma$ of Eq.~(\ref{eq:gamma(omega,omega-prime)}).

\mar{As for the case of the partial secular approximation, the user has the possibility of generating a local unified master equation by passing the bare system Hamiltonian to the code. The frequency clustering is then applied with respect to the eigenvalues of this Hamiltonian.}

\begin{algorithm}[t]
  \caption{Frequency clustering}\label{alg:freq-clustering}
  \begin{algorithmic}
    \Require $\Omega$ a list of Bohr frequencies $\omega$ of length $N$; \newline$w > 0$ the maximum gap between frequencies

    \State Sort $\Omega$
    \State Clusters $\gets [\ ]$
    \State $C \gets [\ ]$
    \Comment{A single cluster}
    \For{$i=0, \dots, N-1$}
    \State $\Delta \omega \gets \Omega[i+1] - \Omega[i]$
    \If{$\Delta \omega < w$}
    \Comment{Frequency $\omega_i\in C$}
    \State Add the frequency $\omega_i$ to the cluster $C$
    \State Continue
    \Else
    \Comment{Frequency $\omega_{i+1}\notin C$}
    \State Add the frequency $\omega_i$ to the cluster $C$
    \State Add $C$ to the collection of all clusters
    \State Start a new cluster: $C \gets [\ ]$
    \EndIf
    \EndFor

    \State Perform the clustering for the last frequency $\omega_N$

    \State \Return Clusters
    
  \end{algorithmic}
\end{algorithm}

\subsection{Block diagonalization}

The symmetries of the Liouvillian superoperator $\mathcal{L}$ can be utilized to transform the matrix representation of $\mathcal{L}$ into a block diagonal form, as explained in Section \ref{sec:symm-cons-quant}. 

The algorithm for obtaining the block diagonal form for the Liouvillian superoperator is detailed in Algorithm \ref{alg:block-diag}. First, the user passes a symmetry operator \mar{$J$} to the algorithm. Then, the algorithm certifies the existence of the symmetry at the superoperator level by checking whether the commutation relation $[\mathcal{L}, \mathcal{J}] = 0$ between the \mar{Liouvillian and the symmetry superoperator $\mathcal{J}=J\otimes\mathbb{1} - \mathbb{1}\otimes J^T$} holds. If it does, then the eigenvectors $\mar{\ket{\omega_i}}$ of $\mathcal{J}$ are calculated. From the eigenvectors a matrix is constructed by appending the column vectors side by side such that the result is a matrix $U$ of the form
\begin{equation}
  \label{eq:trafo-U}
  U = \left[\ket{w_1}, \ket{w_2}, \hdots, \ket{w_{\text{dim}(\mathcal{J})}}\right]\,,
\end{equation}
where $\text{dim}(\mathcal{J}) = \mar{d^2\times d^2}$ if the dimension of the original Hilbert space operator $\text{dim}(J) = \mar{d\times d}$. Then the Liouvillian superoperator can be transformed into a block diagonal form via
\begin{equation}
  \label{eq:blocky-Liouv}
  \mathcal{L}_\text{block} = U^\dagger\mathcal{L}U.
\end{equation}

\begin{algorithm}
  \caption{Block diagonalization}\label{alg:block-diag}
  \begin{algorithmic}
    \Require $\mathcal{L}$ the Liouvillian superoperator;\newline $\mathcal{J}$ the supposed symmetry superoperator

    \State $C \gets \mathcal{L}\mathcal{J} - \mathcal{J}\mathcal{L}$
    \Comment{Check if $[\mathcal{L}, \mathcal{J}] = 0$}
    \If{$C\neq0$}
    \State \Return
    \EndIf

    \State Compute eigenvectors $\ket{w_i}$ of $\mathcal{J}$
    \State $U \gets [\ket{w_1}, \ket{w_2},\hdots,\ket{w_{\dim(\mathcal{J})}}]$
    \Comment{Build $U$}
    \State \Return $U$
  \end{algorithmic}
\end{algorithm}

\section{Examples} \label{sec:examples}
\subsection{Introducing the qubit-resonator chain}\label{sec:intr-qubit-reson}

\begin{figure}
  \centering
  \resizebox{\linewidth}{!}{
    \begin{circuitikz}[cute inductors]
      \tikzstyle{every node}=[font=\large]
      
      \draw (0,0) node[ground] {} to[R] (2,0)
      to[C] node[above]{} (4,0)
      to[C] node[above]{} (7,0)
      to[C] node[above]{} (10,0)
      to[C] node[above]{} (13,0)
      to[C] (15,0) to[R] (17,0) node[ground] {};

      \draw (4,0) -- (4,-1) -- (5,-1)
      to[L] (5,-3) -- (4,-3) node[ground] {} -- (3,-3)
      to[C] (3,-1) -- (4,-1);

      \draw (7,0) -- (7,-1) -- (8,-1) -- (8,-1.5) -- (8.5,-1.5) -- (8.5,-2.5) -- (7.5,-2.5) -- (7.5,-1.5) -- (8.5,-1.5);
      \draw (8.5,-1.5) -- (7.5,-2.5);
      \draw (7.5,-1.5) -- (8.5,-2.5);
      \draw (8,-2.5) -- (8,-3) -- (7,-3) node[ground] {} -- (6,-3) to[C] (6,-1) -- (7,-1);

      \draw (10,0) -- (10,-1) -- (11,-1) -- (11,-1.5) -- (11.5,-1.5) -- (11.5,-2.5) -- (10.5,-2.5) -- (10.5,-1.5) -- (11.5,-1.5);
      \draw (11.5,-1.5) -- (10.5,-2.5);
      \draw (10.5,-1.5) -- (11.5,-2.5);
      \draw (11,-2.5) -- (11,-3) -- (10,-3) node[ground] {} -- (9,-3) to[C] (9,-1) -- (10,-1);

      \draw (13,0) -- (13,-1) -- (14,-1)
      to[L] (14,-3) -- (13,-3) node[ground] {} -- (12,-3)
      to[C] (12,-1) -- (13,-1);

    \end{circuitikz}
  }
  \caption{Circuit diagram of the quantum system composed of two qubits and two resonators with coupling two resistors on either side.}
  \label{fig:circuit-diagram}
\end{figure}
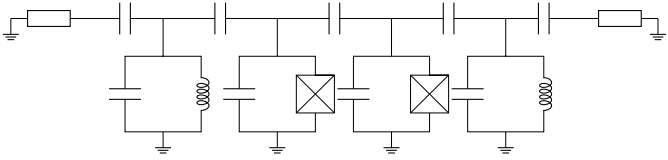

\begin{figure}[t]
  \centering
  \includegraphics[width=\linewidth]{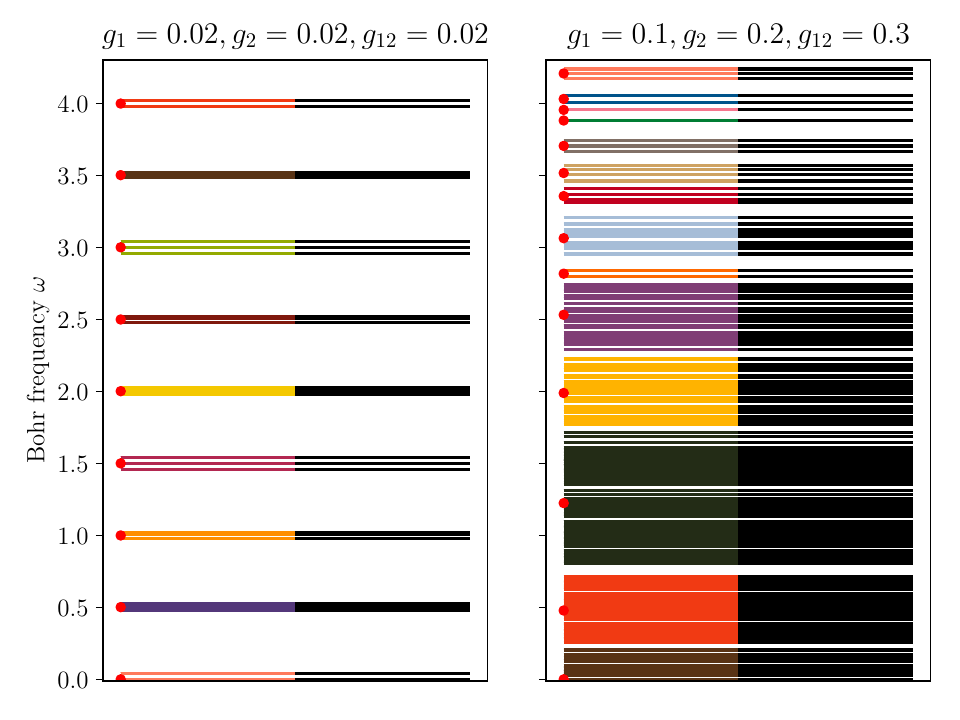}
  \caption{Bohr-frequency spectra for the qubit resonator chain showing the clustering of the frequencies. Each line \mar{represents} a distinct Bohr frequency, the left half of which has been colored to indicate into which cluster Algorithm \ref{alg:freq-clustering} assigns it to. The red dots on the left side of each cluster are the cluster averages $\bar{\omega}$ that we take to represent each cluster. In both cases the boson truncation is $N=4$, $w=0.05$, the qubit frequencies are $\omega_1 = \omega_2 = 1$ and the boson frequencies are $\Omega_\text{L} = \Omega_\text{R} = 1.5$. The coupling varies between the two figures, showing that for weak coupling (on the left) the clustering is clear while for strong coupling (on the right) the clusters cannot be well distinguished from each other. (For interpretation of the references to colour in this figure legend, the reader is referred to the web version of this article.)}
  \label{fig:Bohr-spectra}
\end{figure}

As a testbed for \mar{our code,} we \mar{investigate the} steady-state heat transport through a quantum system \mar{composed of two qubits and two bosonic modes} coupled to two baths at different temperatures. This kind of setup is \mar{of interest for different studies in} quantum thermodynamics \cite{Pekola2015,Goold2016,Vinjanampathy2016,Potts2024}. Experimental realizations \mar{of similar schemes}, where understanding the heat flow through the central quantum system is essential, include for example quantum heat switches \cite{Karimi2017}, valves \cite{Ronzani2018}, and rectifiers \cite{Senior2020} as well as different quantum thermal machines \cite{Tan2017,Aamir2025,Cangemi2024,Bhattacharjee2021}. \mar{Several of these experiments are based on} circuit quantum electrodynamics \cite{Blais2004,Blais2021}, \mar{where resonators play the role of the bosonic modes, which can be coupled to superconducting qubits.} 

\mar{In our example, we consider a 1D chain composed of two central superconducting qubits, each of them coupled to a distinct resonator, forming a symmetric structure. In turn, each resonator is interacting with a thermal bath.} This geometry is motivated by the study of photonic heat transport in a quantum heat valve \cite{Ronzani2018}, \mar{whose experimental implementation adopted the same setup, with the only difference that a single central qubit was used, instead of two. There,} the magnitude of heat current between the two baths was controlled by varying the resonance frequency of the central qubit in the chain. A similar system has also been studied in \cite{Satrya2023} using completely classical methods of electromagnetic field simulation and circuit engineering. Fig.~\ref{fig:circuit-diagram} shows the circuit diagram of \mar{the setup under consideration}, where we see that \mar{role of} the baths \mar{is played by two} resistors. \mar{The bosonic modes are} parallel LC-oscillators, \mar{while the} qubits are Josephson junctions shunted by a capacitor, in a so called transmon design \cite{Koch2007}. Fig.~\ref{fig:schematic-figure-of-the-studied-Hamiltonian} shows an abstract representation of the system \mar{as a 1D chain}.

The Hamiltonian of the circuit in Fig.~\ref{fig:circuit-diagram} can be derived using the toolbox of circuit quantum electrodynamics \cite{Blais2021,Krantz2019,Rasmussen2021,Vaaranta2022,Cattaneo2021}. \mar{Under the standard rotating wave approximation, it reads} 
\begin{equation}
  \label{eq:Hamiltonian-system-RWA}
  \begin{split}
    H_\text{system} = &\frac{1}{2}\omega_1\sigma_z^{(1)} + \Omega_\text{L}a_\text{L}^\dagger a_\text{L} +  g_1(\sigma_+^{(1)}a_\text{L} + \sigma_-^{(1)}a_\text{L}^\dagger) \\
    + &\frac{1}{2}\omega_2\sigma_z^{(2)} + \Omega_\text{R}a_\text{R}^\dagger a_\text{R} +  g_2(\sigma_+^{(2)}a_\text{R} + \sigma_-^{(2)}a_\text{R}^\dagger) \\
    + & g_{12}(\sigma_+^{(1)}\sigma_-^{(1)} + \sigma_-^{(1)}\sigma_+^{(2)})\,,
  \end{split}
\end{equation}
where $\omega_1$ and $\omega_2$ are the frequencies of the two qubits, $\Omega_\text{L}$ and $\Omega_\text{R}$ are the frequencies of the left and right resonators respectively, while $g_1, g_2$ and $g_{12}$ are the couplings between the left qubit and resonator, right qubit and resonator, and the qubits respectively. The Hamiltonian is of the form of a double Jaynes-Cummings model \cite{Shore1993,Larson2021,Jaynes1963}, with an additional qubit-qubit interaction. Fig.~\ref{fig:Bohr-spectra} shows the Bohr frequency spectrum of the Hamiltonian and indicates that the frequencies can be well separated into distinct clusters for small couplings $g,g_{12}$ whereas for stronger couplings the clustering fails. \new{However, for very strong couplings with $g\sim g_{12} > 1$ well-defined clusters reappear.}

The resistors on either edge of the circuit in Fig.~\ref{fig:circuit-diagram} are described using the Caldeira-Leggett model, and their Hamiltonians are given by \cite{Cattaneo2021}
\begin{equation}
  \label{eq:Baths-Ham}
  H_{\text{Bath}, i} = \sum_{k=0}^\infty\Omega_{k, i}b_{k,i}^\dagger b_{k,i}\,,
\end{equation}
where $\Omega_k$ is the resonance frequency of mode $k$ and $i$ refers to the left or right bath. The \mar{characterization of the} bosonic baths and some of the properties that follow from it are explained in more detail in \ref{sec:coupl-boson-bath}. In what follows, \mar{we will make use of these properties without mentioning them explicitly.}

\begin{figure*}
  \centering
  \resizebox{0.75\linewidth}{!}{
    \begin{tikzpicture}
      \node[cloud, draw, aspect=2.7, cloud puffs=20] (c1) at (0,0) {\small$\sum_k\Omega_k b_k^\dagger b_k$};
      \node[circle, draw, minimum width=1.25cm, thick] (r1) at (3,0) {};
      \node[rectangle, draw, minimum width=1cm, minimum height=1cm, thick] (q1) at (5,0) {};
      \node[rectangle, draw, minimum width=1cm, minimum height=1cm, thick] (q2) at (7,0) {};
      \node[circle, draw, minimum width=1.25cm, thick] (r2) at (9,0) {};
      \node[cloud, draw, aspect=2.7, cloud puffs=20] (c2) at (12,0) {\small$\sum_k\Omega_k b_k^\dagger b_k$};

      \draw[<->, thick] (c1.east) to [out=30, in=150] node [above, yshift=3mm] {} (r1.west);
      \draw[<->, thick] (r1.east) to [out=30, in=150] node [above, yshift=3mm] {$g_1$} (q1.west);
      \draw[<->, thick] (q1.east) to [out=30, in=150] node [above, yshift=3mm] {$g_{12}$} (q2.west);
      \draw[<->, thick] (q2.east) to [out=30, in=150] node [above, yshift=3mm] {$g_2$} (r2.west);
      \draw[<->, thick] (r2.east) to [out=30, in=150] node [above, yshift=3mm] {} (c2.west);

      \begin{scope}[scale=0.11, xshift=27.5cm, yshift=-3.5cm]
        \draw[domain=-2.6:2.6, smooth, variable=\x] plot ({\x}, {\x*\x});
        \draw[domain=-1.41:1.41, variable=\x] plot ({\x}, {2});
        \draw[domain=-2:2, variable=\x] plot ({\x}, {4});
        \draw[domain=-2.45:2.45, variable=\x] plot ({\x}, {6});
      \end{scope}

      \begin{scope}[xshift=5cm]
        \draw[domain=-0.4:0.4, variable=\x] plot ({\x}, {-0.25});
        \draw[domain=-0.4:0.4, variable=\x] plot ({\x}, {0.25});
        \draw[<->] (-0.3,0.25) -- (-0.3,-0.25);
        \node at (0.1,0) {$\omega_1$};
      \end{scope}

      \begin{scope}[xshift=7cm]
        \draw[domain=-0.4:0.4, variable=\x] plot ({\x}, {-0.25});
        \draw[domain=-0.4:0.4, variable=\x] plot ({\x}, {0.25});
        \draw[<->] (-0.3,0.25) -- (-0.3,-0.25);
        \node at (0.1,0) {$\omega_2$};
      \end{scope}

      \begin{scope}[scale=0.11, xshift=82cm, yshift=-3.5cm]
        \draw[domain=-2.6:2.6, smooth, variable=\x] plot ({\x}, {\x*\x});
        \draw[domain=-1.41:1.41, variable=\x] plot ({\x}, {2});
        \draw[domain=-2:2, variable=\x] plot ({\x}, {4});
        \draw[domain=-2.45:2.45, variable=\x] plot ({\x}, {6});
      \end{scope}
      
    \end{tikzpicture}
  }
  \caption{Pictorial representation of the nearest neighbour coupling between the different elements of the qubit-resonator chain.}
  \label{fig:schematic-figure-of-the-studied-Hamiltonian}
\end{figure*}

\begin{figure*}
  \centering
  \begin{subfigure}{0.49\textwidth}
    \centering
    \includegraphics[width=\linewidth]{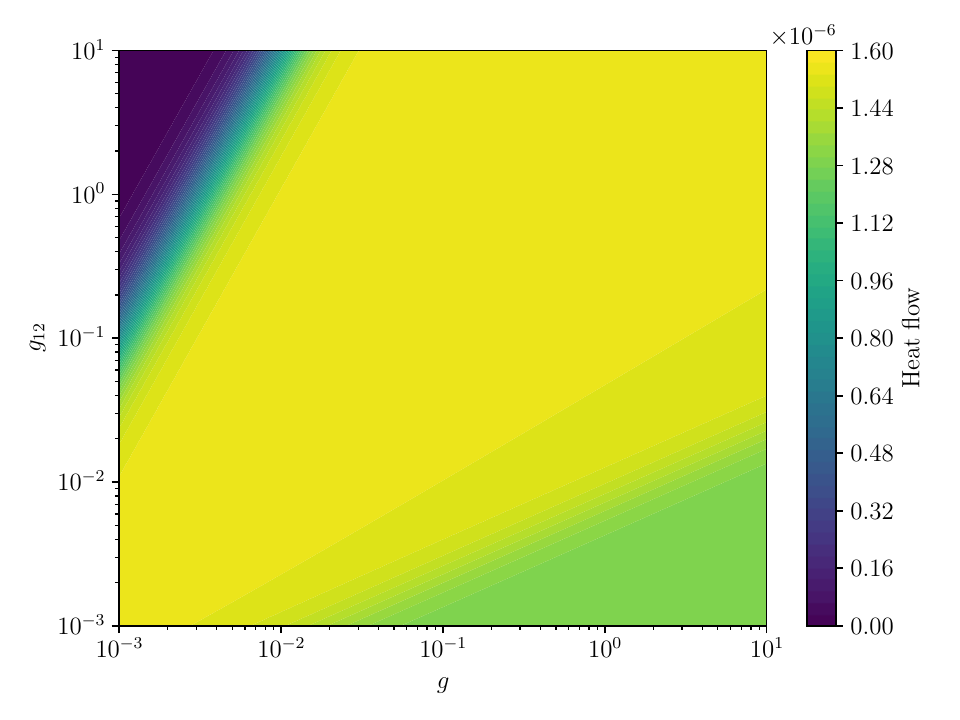}
    \caption{$\Omega_\text{L} = \Omega_\text{R} = 1$ and $T_\text{L} = 0.5, T_\text{R} = 0.1$. We observe strong steady heat flow through the system for most of the coupling parameter space, even for small couplings due to full resonance between the resonators and the qubits.}
    \label{fig:resonant-heatflow-local}
  \end{subfigure}
  \hfill
  \begin{subfigure}{0.49\textwidth}
    \centering
    \includegraphics[width=\linewidth]{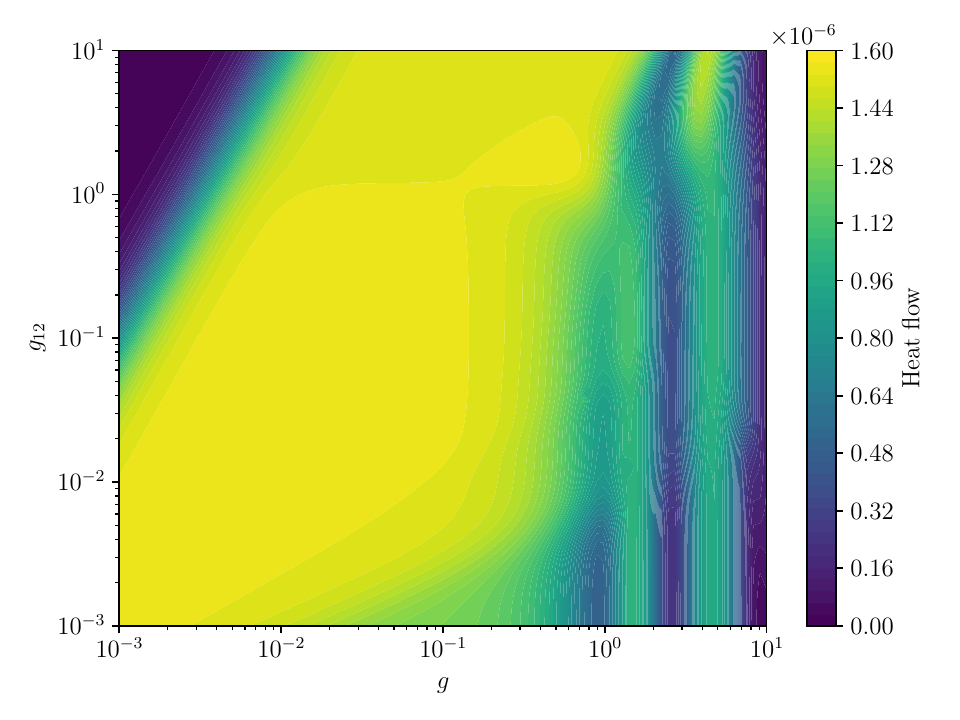}
    \caption{$\Omega_\text{L} = \Omega_\text{R} = 1$ and $T_\text{L} = 0.5, T_\text{R} = 0.1$. The heat flows coincide for $g\lesssim 0.1$ with the local master equation predictions, while for $g > 0.1$ the global master equation predicts a more complex structure of the heat flow in the coupling parameter space.}
    \label{fig:resonant-heatflow-global}
  \end{subfigure}
  \hfill
  \begin{subfigure}{0.49\textwidth}
    \centering
    \includegraphics[width=\linewidth]{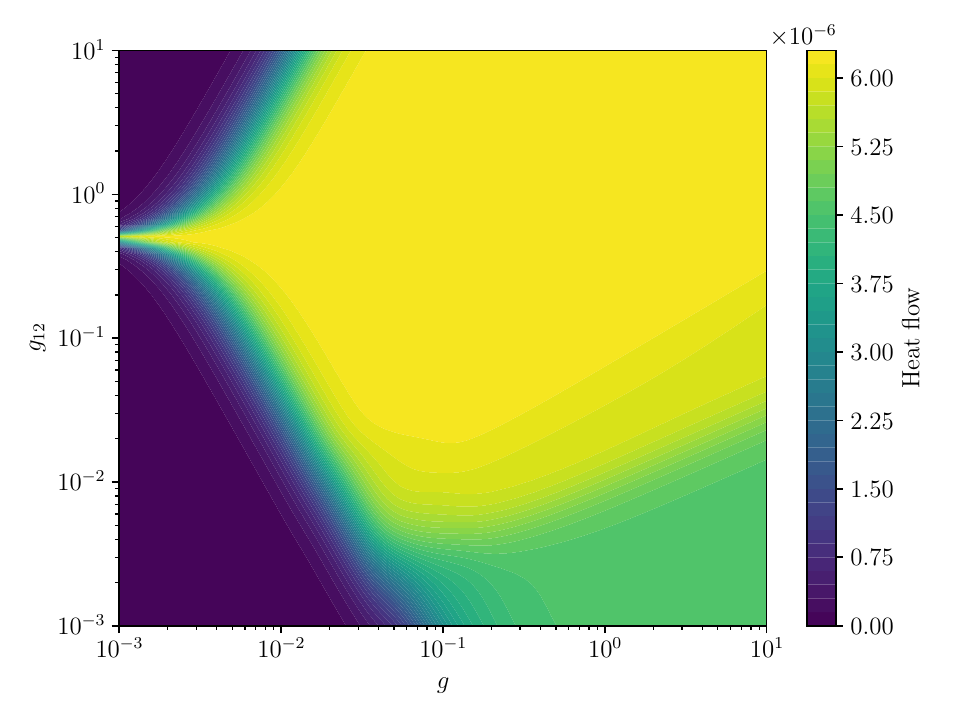}
    \caption{$\Omega_\text{L} = \Omega_\text{R} = 1.5$ and $T_\text{L} = 1, T_\text{R} = 0.1$. A sweet spot of the qubit-qubit coupling $g_{12}$ is visible, allowing for a strong heat flow through the system although the coupling $g$ is weak.}
    \label{fig:non-resonant-heatflow-local}
  \end{subfigure}
  \hfill
  \begin{subfigure}{0.49\textwidth}
    \centering
    \includegraphics[width=\linewidth]{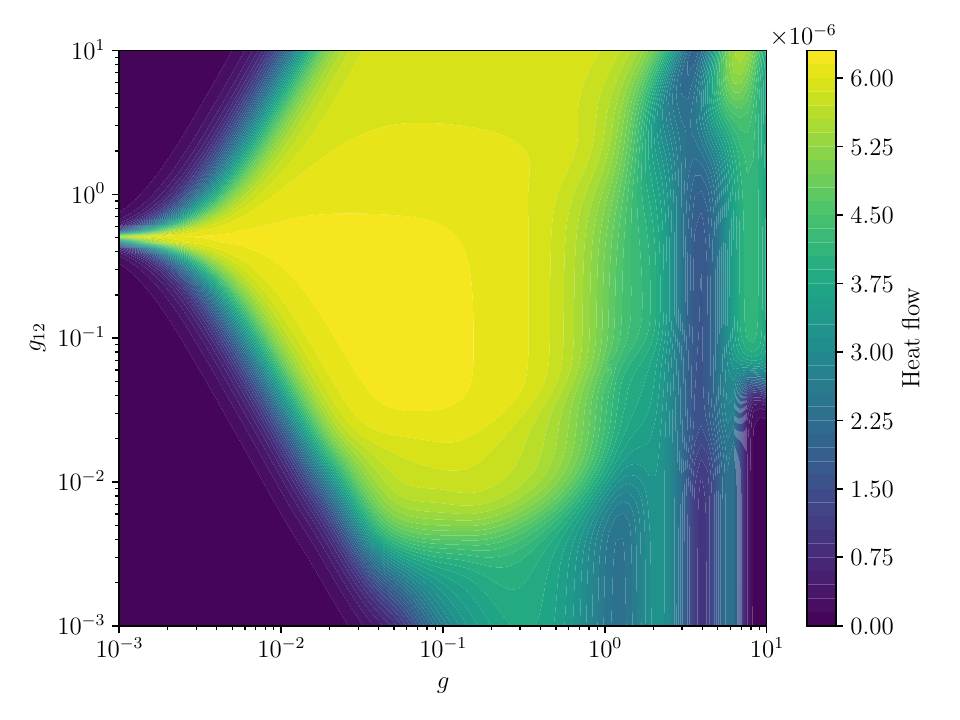}
    \caption{$\Omega_\text{L} = \Omega_\text{R} = 1.5$ and $T_\text{L} = 1, T_\text{R} = 0.1$. The heat flows coincide for $g\lesssim 0.1$ with the local master equation predictions, while for $g > 0.1$ the global master equation predicts a more complex structure of the heat flow in the coupling parameter space.}
    \label{fig:non-resonant-heatflow-global}
  \end{subfigure}
  \caption{\new{Steady-state heat flow through the qubit-resonator chain as a function of the qubit-qubit coupling $g_{12}$ and the qubit-resonator coupling $g$. Top row: all qubits and resonators are at resonance. Bottom row: the resonators are detuned from the qubits. The left column: solutions obtained via local master equations. The right column: solutions obtained via global master equations with partial secular approximation. Other parameters are $\omega_1 = \omega_2 = 1$, $\alpha_\text{L} = \alpha_\text{R} = 0.01$, $\chi_\text{L} = \chi_\text{R} = 0.1$, $\omega_\text{c} = 100$. The resonator Hilbert space has been truncated to $N=4$ levels. The partial secular approximation cut-off value has been set to $C_\text{PSA} = 10^4$ for global master equations while for the local ones we can safely employ the full secular approximation (so $C_\text{PSA} = 0$) due to the baths being separate.}}
  \label{fig:heatflow-collection}
\end{figure*}

\subsection{Steady-state heat flow through the system}

Our aim is to study the steady-state heat flow between the two baths connected to the qubit-resonator chain. \mar{With} that in mind, we derive the master equation in both global and local regimes with full and partial secular approximation. While both master equations inherently assume weak coupling to the environment, the local master equation assumes additionally that the inner system couplings are also weak. Therefore, the dissipators of Eq. (\ref{eq:Redfield-ME-Schröd-pic}) deal with jump operators that act only on the subsystem connected to the baths, which  in our case consists of the  resonators. In this case, the master equation can be fully written down due to its relatively simple form:
\begin{equation}
  \label{eq:Local-ME-Lindblad}
  \frac{\d}{\d t}\rho_\text{S} =
  \begin{aligned}[t]
    &-\i\left[H_\text{S} + H_\text{LS}, \rho_\text{S}\right] \\
    &+ \gamma_\text{L}(1 + \bar{n}(\Omega_\text{L}))\left(a_\text{L}\rho_\text{S}a_\text{L}^\dagger - \frac{1}{2}\left\{a_\text{L}^\dagger a_\text{L}, \rho_\text{S}\right\}\right) \\
    &+  \gamma_\text{L}\bar{n}(\Omega_\text{L}) \left(a_\text{L}^\dagger\rho_\text{S}a_\text{L} - \frac{1}{2}\left\{a_\text{L}a_\text{L}^\dagger, \rho_\text{S}\right\}\right) \\
    &+ \gamma_\text{R}(1 + \bar{n}(\Omega_\text{R}))\left(a_\text{R}\rho_\text{S}a_\text{R}^\dagger - \frac{1}{2}\left\{a_\text{R}^\dagger a_\text{R}, \rho_\text{S}\right\}\right) \\
    &+ \gamma_\text{R} \bar{n}(\Omega_\text{R}) \left(a_\text{R}^\dagger\rho_\text{S}a_\text{R} - \frac{1}{2}\left\{a_\text{R}a_\text{R}^\dagger, \rho_\text{S}\right\}\right) \,,
  \end{aligned}
\end{equation}
where
\begin{equation}
  \gamma_i = 2\pi\alpha_i^2J_i(\Omega_i)
\end{equation}
is the coefficient describing the coupling to the baths from either left ($i=\text{L}$) or right ($i=\text{R}$) with spectral density of the bath $J(\Omega)$ \cite{breuer2002theory,Cattaneo2021} and dimensionless coupling $\alpha_i$. Additionally, $\bar{n}(\Omega)$ denotes the Bose-Einstein distribution
\begin{equation}
  \label{eq:BE-dist}
  \bar{n}(\Omega) = \frac{1}{\e^{\Omega/T} - 1}\,,
\end{equation}
with temperature $T$ of the corresponding bath. Note that the form of the local master equation remains the same in both full and secular approximation due to the baths being separate. For common baths, the local master equation would also change \mar{depending on whether the full or partial secular approximation is performed} \cite{Cattaneo2019}.

The \mar{structure of the} global master equation is\mar{, instead}, much more complex, as \mar{in this case} the jump operators act globally on the whole system. Therefore, it can be written in its most general form as in Eq.~(\ref{eq:Redfield-ME-Schröd-pic}), where the jump operators $A(\omega)$ act on the quantum system as a whole. 

The steady-state heat flow is \mar{defined as}
\begin{equation}
  \label{eq:steadystate-heatflow}
  J_i = \Tr\left[H_\text{S}\mathcal{D}_i[\rho_\infty]\right]\,,
\end{equation}
where the dissipator $\mathcal{D}_i$ is either from the left ($i=\text{L}$)  or right ($i=\text{R}$) bath. Note that, since $\rho_\infty$ is the steady state, we have $J_\text{L} + J_\text{R} = 0$.

\begin{figure}[t]
  \centering
  \includegraphics[width=\linewidth]{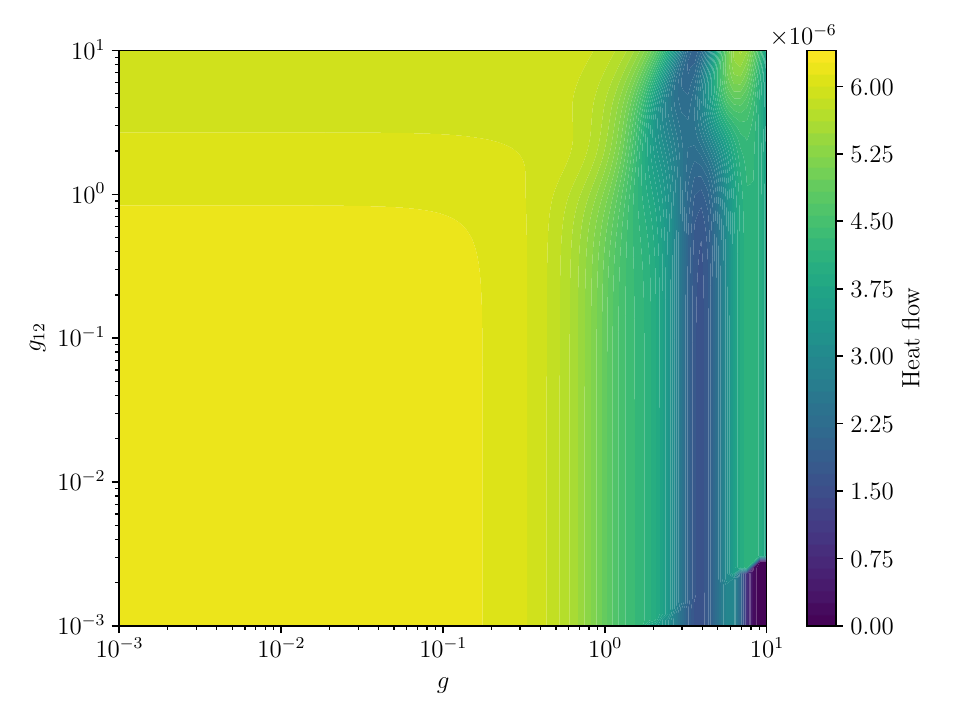}
  \caption{Heat flow through the qubit-resonator chain with $\Omega_\text{L} = \Omega_\text{R} = 1.5$ and $T_\text{L} = 1, T_\text{R} = 0.1$ calculated using the global master equation with the full secular approximation. The steady-state heat flow at small couplings $g$ is greatly overestimated due to breaking of the full secular approximation.}
  \label{fig:global-FSA}
\end{figure}

\begin{figure}[t]
  \centering
  \includegraphics[width=\linewidth]{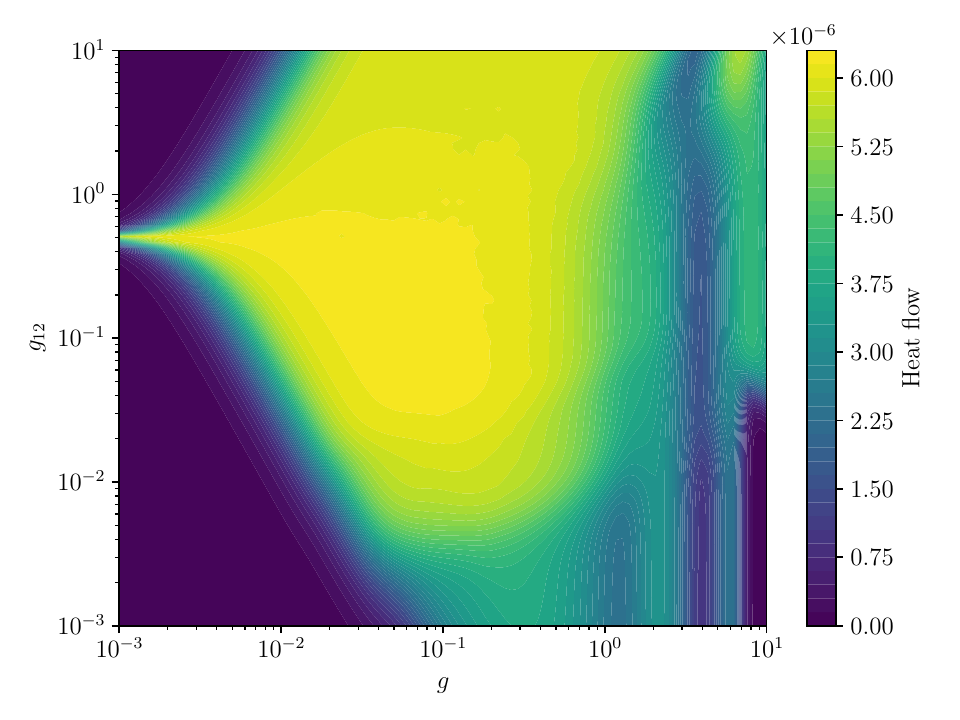}
  \caption{Heat flow through the qubit-resonator chain with $\Omega_\text{L} = \Omega_\text{R} = 1.5$ calculated using the unified master equation approach with cluster width $w = 0.01$. \new{The unified master equation approach matches well with the global master equation with partial secular approximation from Fig.~\ref{fig:non-resonant-heatflow-global}, while being significantly faster to compute numerically.}}
  \label{fig:unified-me-res-heatflow}
\end{figure}


We compare the effect of partial secular and full secular approximations with global and local master equations on the heat flow through the system, as well as the effect of the qubit-resonator frequency detuning, while changing the coupling coefficients between the qubits and resonators. For simplicity, we set the qubit-resonator couplings on either side to be equal, $g_1 = g_2 = g$, while the qubit-qubit coupling $g_{12}$ can be changed independently. The results \mar{are shown} in Figs.~\ref{fig:heatflow-collection} --~\ref{fig:global-FSA}. In all figures, the qubit frequencies have been normalized to unity ($\omega_1 = \omega_2 = 1$), the couplings strength of the system-bath interaction has been set to $\alpha_\text{L} = \alpha_\text{R} = 0.01$ on either side of the system, and the baths are \mar{coupled to the system through an} Ohmic spectral density with Drude cutoff [see Eq.~(\ref{eq:spectral-dens}) in \ref{sec:coupl-boson-bath}] with coefficient $\chi_\text{L} = \chi_\text{R} = 0.1$ and cutoff frequency $\omega_\text{c} = 100$.\footnote{In the code the spectral density can be arbitrary and defined by the user. The only restriction is that the three arguments, angular frequency $\omega$, parameter $\chi$ and the cut-off frequency $\omega_\text{c}$, need to be present in this order.} The Hilbert space dimension \mar{for both resonators} is truncated to $N=4$, which corresponds to highest excited state population in the thermal state being less than $10^{-2}$ \new{(this was chosen to obtain reasonable computation times)}. Therefore the dimensionality of the whole Hilbert space is $2^2\cdot4^2 = 64$. The white \mar{areas} in some of the figures correspond to negative values of the heat flow, calculated from the hot to cold bath, which have been changed to NaNs as they are unphysical. These values arise from numerical instabilities for large qubit-resonator couplings.

Figs.~\ref{fig:resonant-heatflow-local} and \ref{fig:resonant-heatflow-global} show the heat flow through the system in the fully resonant case where the boson frequencies have been set to that of the qubits, i.e. we have $\Omega_\text{L} = \omega_1 = \omega_2 = \Omega_\text{R} = 1$ and the temperatures of the baths are $T_\text{L} = 0.5$ and $T_\text{R} = 0.1$. We can see how the heat flow given by the local master equation, in Fig.~\ref{fig:resonant-heatflow-local}, matches the one given by the global master equation, in Fig.~\ref{fig:resonant-heatflow-global}, in the regime of \mar{weak qubit-resonator coupling,} $g\lesssim 0.1$. Thus, the local master equation \mar{reproduces} the results of the global one.

In the strong qubit-resonator coupling regime, where $g > 0.1$, we can see clear differences between the heat flows given by the global and local master equations. The local one predicts strong steady-state heat flow for all values of the coupling except for very weak qubit-qubit coupling $g_{12}$, where the flow starts to weaken. However, the global master equation shows a complex structure in the heat flow when the couplings are varied in the strong coupling regime, showing that at some values of the qubit-resonator coupling $g$ the heat flow is strongly inhibited whereas increasing or decreasing the coupling allows heat to flow through the system once more. 

Figs.~\ref{fig:non-resonant-heatflow-local} and \ref{fig:non-resonant-heatflow-global} show the heat flow in the nonresonant case where the resonators have been detuned \mar{from} the qubit  frequencies, as $\Omega_\text{L} = \Omega_\text{R} = 1.5$. The bath temperatures are $T_\text{L} = 1$, $T_\text{R} = 0.1$. In this regime, some \mar{interesting structures of the heat flow in the parameter space arise even when using the local master equation} (see Fig.~\ref{fig:non-resonant-heatflow-local}). We \mar{observe} that, at very small values of the qubit-resonator coupling $g$, there exists a narrow region of values of the qubit-qubit coupling $g_{12}$ \mar{where heat flows} the system as \mar{strongly as} in the strong coupling regime. This can be explained by noting that at very small values of $g$ the qubit-qubit system can be viewed as an effective four-level system with a Hamiltonian
\begin{equation}
  \label{eq:qq-Ham}
  H_{\text{qq}} = \frac{1}{2}\omega_1\sigma_z^{(1)} + \frac{1}{2}\omega_2\sigma_z^{(2)} + g_{12}(\sigma_+^{(1)}\sigma_-^{(2)} + \sigma_-^{(1)}\sigma_+^{(2)})\,.
\end{equation}
When setting $\omega_1 = \omega_2 = 1$, the eigenenergies of this Hamiltonian are given by $\{\pm1, \pm g_{12}\}$. Therefore we can find resonance between the Bohr frequency $1+g_{12}$ and $\Omega_\text{L/R}$ by choosing $g_{12} = \Omega_\text{L/R} - 1$. If $\Omega_{\text{L/R}} = 1.5$ as in our examples, then we should observe incresed heat flow for $g_{12} = 0.5$, which we do. The steady-state heat flow being maximal around when there exist resonant transitions between the qubit-qubit and resonator subsystems is consistent with experimental findings of \cite{Ronzani2018}.

In the global case in Fig.~\ref{fig:non-resonant-heatflow-global} the same preferred range of $g_{12}$ values is visible. Additionally, more complex structures \mar{arise} for larger values of coupling $g$. These structures are similar to \mar{those} in Fig.~\ref{fig:resonant-heatflow-global}, in the case of a fully resonant system studied using the global master equation. Note that, \mar{once again}, the global and local master equations agree for weak qubit-resonator couplings $g\lesssim0.1$.

\new{As pointed out earlier in Section \ref{sec:unif-mast-equat}, the partial secular approximation suffers from issues related to non-positivity of the master equation. For the sake of numerical simulations, these negativities should anyway be small if the Born-Markov approximations are sound \cite{Hartmann2020}. }

\begin{figure}
  \centering
  \includegraphics[width=\linewidth]{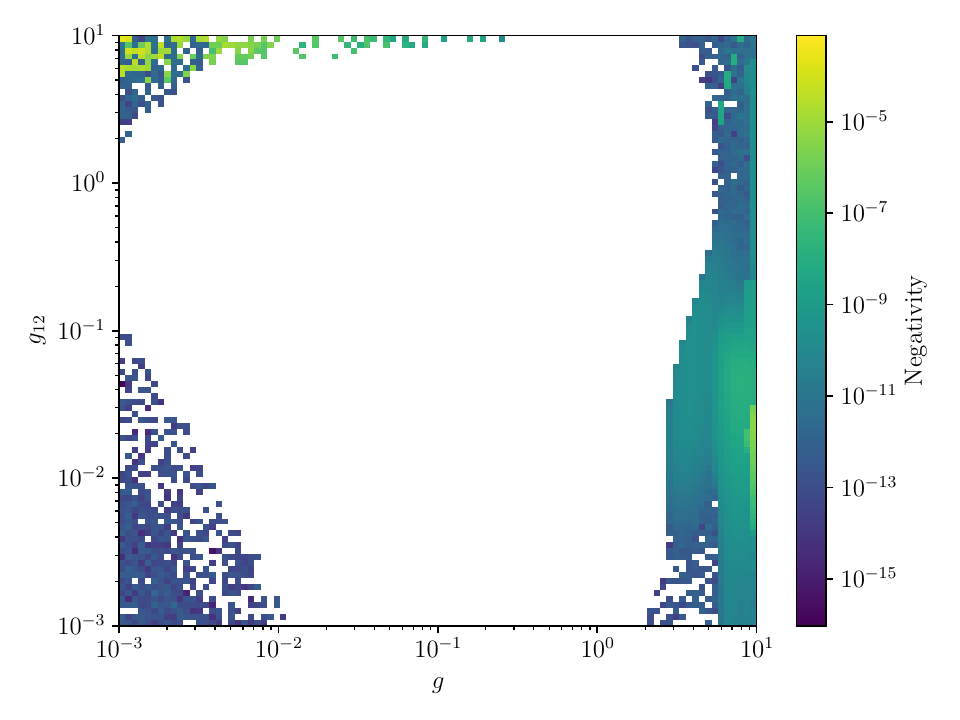}
  \caption{\new{Negativity of the steady-state density matrix generated by a global master equation with partial secular approximation and $\Omega_\text{L} = \Omega_\text{R} = 1.5$. All parameters are set as in Fig.~\ref{fig:non-resonant-heatflow-global}. White spots correspond to negativity equal to zero.
  }}
  \label{fig:negativity-plot}
\end{figure}

\new{To check this is the case in our example, we need to understand the severity of non-positivity of our numerical solutions. Hence, we compute the negativity of the steady-state density matrix $\rho_\infty$ by taking the absolute value of the sum of its negative eigenvalues, through a subroutine available in our code \cite{Vaaranta2025_GitHub}. The negativity of the steady-state solutions for the case of Fig.~\ref{fig:non-resonant-heatflow-global} is plotted in Fig.~\ref{fig:negativity-plot}. We can see that mostly the negativity is zero (the white regions in the plot) or has a negligible value. Therefore, we argue that although the partial secular approximation might suffer from non-positivity issues, the negativity of the solutions is negligible in the vast majority of the cases and should therefore not discourage us from using the partial secular approximation. Moreover, small negativity values may also be due to numerical errors in the computation of the steady state. We indeed observed that these small values also occur in the steady state of the full secular equation, which is completely positive by construction, and can be traced to (for all practical purposes negligible) numerical instabilities in the diagonalization of the Liouvillian.
}

\mar{Next,} in Fig.~\ref{fig:global-FSA} we \mar{show} the steady-state heat flow in the non-resonant case $\Omega_\text{L} = \Omega_\text{R} = 1.5$ \mar{computed} using the full secular approximation \mar{instead of the partial one} in the global master equation. There, we notice clear discrepancies in the heat flow predictions, when compared to Fig.~\ref{fig:non-resonant-heatflow-global}. The full secular approximation overestimates the steady-state heat flows everywhere in the weak coupling regime, where \mar{it is known to break} down, as also observed in \cite{Cattaneo2019, Hofer2017,Gonzalez2017}. In the ultra-strong coupling regime (with $g \gtrsim 3$) the results \mar{get closer to those of} the partial secular approximation, as \mar{it can be shown that} many quasidegeneracies in the Bohr frequency spectrum are lifted as the spectrum regains a gapped form. 

Lastly, in Fig.~\ref{fig:unified-me-res-heatflow} we show 
the steady-state heat flow 
using the unified master equation 
with the same parameters as in Fig.~\ref{fig:non-resonant-heatflow-global}. We notice that the results match, showing that the unified master equation is able to capture the same \mar{quantitative physical} behaviour as the term-by-term partial secular approximation\new{, for a sufficiently small parameter $w$. Indeed, in Fig.~\ref{fig:unified-me-res-heatflow} the maximum separation between the Bohr frequencies was chosen as $w=0.01$ in order to prevent inaccuracies from arising near strong couplings $g\sim g_{12}\sim 0.1$, as shown in Fig.~\ref{fig:Bohr-spectra}, where $w=0.05$ was chosen instead. Using $w=0.01$, we obtain many more narrower clusters that are separated by relatively small gaps.} 

\new{While, in principle, the unified equation is neglecting some non-secular transitions between nearby clusters in this scenario, in our numerical results this has no relevant effects. We may interpret this as an indication that the failure of the full secular approximation is mostly due to the neglection of all non-secular cross-terms between nearby frequencies, while the unified equation neglects only some of them, namely those between two adjacent clusters with spacing larger than 0.01. It is also noteworthy that at very strong coupling $g\sim g_{12} > 1$ the clustering becomes again well defined, like in the case of weak coupling}. 

Moreover, we \mar{have noticed} that, by using the \mar{unified master equation}, the computation times are significantly faster than \mar{in the case of} the partial secular approximation. Solving \mar{for} the steady-state heat flow for the same system parameters for a single point took approximately \new{16} seconds with \mar{unified equation}, while it took approximately \new{470} seconds \mar{for the partial secular approximation}, \new{when computing the point with $g=g_{12}=0.001$ in Fig.~\ref{fig:non-resonant-heatflow-global}}. 

\mar{Computing the} unified master equation is more time efficient \mar{than the partial secular approximation} because, \mar{in the latter, the code needs to check whether to keep or discard} each cross term separately. \mar{In contrast, in the unified master equation this check is not necessary, and the only additional step consists of creating the clusters.} 
The \mar{Bohr frequencies} within the same cluster are then represented by \mar{a single average frequency, leading to a single} jump operator $A(\bar{\omega})$. \mar{Therefore,} the number of jump operators \mar{in the master equation} decreases drastically, which \mar{implies a shorter computational time}.

\section{Conclusions} \label{sec:conclusions}

In this manuscript, we have \mar{presented a code for the numerical implementation of the partial secular approximation and unified master equation for structured open quantum systems}. \mar{These approaches generalize the standard full secular approximation to regimes where the latter is not valid, namely when the system has some quasidegenerate Bohr frequencies, as in the case of weakly coupled multipartite open quantum systems. The code can implement both the local and global master equation.} 


\mar{We have tested the code} by computing the steady-state heat flow through a symmetric qubit-resonator system \mar{coupled to two heat baths with different temperatures at the edges}. Our results show an agreement between the \mar{partial secular approach and the unified master equation.} 
Moreover, we have compared the local and global implementations of the master equation, \mar{showing that, as expected, they predict the same value} of the steady-state heat flow for small qubit-resonator couplings 
We have also noted the breakdown of the full secular approximation  for small couplings, \mar{where the steady-state heat flow is greatly overestimated}. This observation is in line with previous literature and shows that our numerical implementation of the secular approximation is able to capture both regimes.

Interestingly, there exists a sweet spot for the qubit-qubit coupling $g_{12}$ within the chain system. Even for very weak qubit-resonator coupling the heat is able to flow \mar{freely} through the system for a suitable choice of $g_{12}$. In this regime the qubit-qubit system can be treated as an effective four-level system, with Bohr frequencies depending on the coupling $g_{12}$. With a suitable choice of this coupling a resonance between the resonators and qubit-qubit system appears, which enhances the heat flow even if the coupling to the resonators is very weak.

The \mar{implementation of the} unified master equation is computationally less demanding than the partial secular approximation. This is due to the clustering of the Bohr frequencies, which significantly reduces the number of necessary matrix multiplications. The jump operators between different clusters are not considered at all, whereas in the term-by-term partial secular approximation all the different jump operator combinations need to be compared. However, the \mar{partial secular approximation} is simpler in its nature, and it works for a wider class of systems than the \mar{the unified master equation, which is valid only when the Bohr frequencies form different well-separated clusters. For instance, for complex many body systems with a dense spectrum of Bohr frequencies, the unified equation may not be applicable. \new{This said, in our numerical example the unified equation captures the same steady-state behavior as the partial secular approximation even for scenarios where the clusters are separated only by small gaps, suggesting that its practical applicability may be broader than expected from theory.}}

\mar{Finally}, we have also discussed the \mar{possible presence of} symmetries of the \mar{open system dynamics} and how they can be employed to block diagonalize the matrix representation of the \mar{Liouvillian}. \mar{Our code} can \mar{take these symmetries into account, allowing for dimensionality reduction}.

\section*{Acknowledgements}

We would like to thank Paolo Muratore-Ginanneschi for insightful comments and for reading the draft. We also would like to thank Jukka Pekola for useful suggestions and discussions. We acknowledge the financial support of the Research Council of Finland through the Finnish Quantum Flagship project (358878, UH) and (35887, Aalto). The authors wish to thank the Finnish Computing Competence Infrastructure (FCCI) for supporting this project with computational and data storage resources. M.C. acknowledges funding from  the Research Council of Finland through the Centre of Excellence program grant 336810 and from COQUSY project PID2022-140506NB-C21 
fund-ed by MCIN/AEI/10.13039/501100011033.

\section*{Author contributions}
A.V. developed the code presented in this work, performed the numerical simulations, and wrote the first draft of the manuscript. M.C. proposed the scientific questions, supervised the project, and revised the draft. Both authors contributed to scientific discussions and to the analysis of the results. 

\appendix
\section{Coupling to a bosonic bath} \label{sec:coupl-boson-bath}

\mar{In this paper, we focus on open quantum systems coupled linearly to bosonic thermal baths. This assumption, apart from being widely employed in the literature, is justified by different experimental scenarios where the bath is bosonic and thermal. For instance, this is the case for a superconducting quantum circuit coupled to a resistor,} which can be modelled as an infinite bath of bosonic modes \mar{at the resistor temperature} via the Caldeira-Leggett model \cite{Vool2017}. More precisely, the bath Hamiltonian is given by
\begin{equation}
  \label{eq:bath-Hamiltonian}
  H_\text{B} = \sum_{k=0}^\infty \Omega_k b_k^\dagger b_k\,,
\end{equation}
where $\Omega_k$ is the frequency of the $k$th mode. The interaction between the system and the bath is mediated by the bath operators 
in the interaction Hamiltonian $H_\text{I}$ [see Eq.~(\ref{eq:H_I-def})] of the total Hamiltonian as
\begin{equation}
  \label{eq:Bath-op-interaction-Ham}
  B = \sum_{k=0}^\infty f_k(b_k^\dagger \pm b_k)\,,
\end{equation}
where $f_k$ describes the coupling strength \mar{of the $k$th mode} to the system. Note that here we perform all the calculations for the case of only a single bath operator in the interaction Hamiltonian, which drops the index $\beta$ from Eq.~(\ref{eq:H_I-def}). Extension to the case of multiple operators is straightforward.

Here we consider coupling to both position ($+$ sign) and momentum ($-$ sign) of the bath modes. In the interaction picture this operator reads
\begin{equation}
  \label{eq:Bath-op-interaction-Ham-int-pic}
  \tilde{B}(t) = \sum_{k=0}^\infty f_k(\e^{\i\Omega_k t}b_k^\dagger \pm \e^{-\i\Omega_k t}b_k)\,.
\end{equation}
With these definitions and taking the state of the bath $\rho_\text{B}$ to be thermal, we can compute the bath correlation function as
\begin{equation}
  \label{eq:bath-corr-func}
  \begin{aligned}
    \left\langle\tilde{B}^\dagger(\tau)\tilde{B}(0)\right\rangle &= \Tr\left[\tilde{B}^\dagger(\tau)\tilde{B}(0)\rho_\text{B}\right] \\
                                                                                      &= \sum_{k,k^\prime}f_{k}^*f_{k^\prime}\e^{-\i\Omega_{k}\tau}
                                                                                        \begin{aligned}[t]
                                                                                          &\Big(\Tr\left\{b_{k}b_{k^\prime}^\dagger\rho_\text{B}\right\} \\
                                                                                          &\pm  \Tr\left\{b_{k}b_{k^\prime}\rho_\text{B}\right\}\Big) 
                                                                                        \end{aligned}
    \\
                                                                                      &+ f_{k}^*f_{k^\prime}\e^{\i\Omega_{k}\tau}
                                                                                        \begin{aligned}[t]
                                                                                          &\Big(\Tr\left\{b_{k}^\dagger b_{k^\prime}\rho_\text{B}\right\} \\
                                                                                               &\pm \Tr\left\{b_{k}^\dagger b_{k^\prime}^\dagger\rho_\text{B}\right\}\Big)\,.
                                                                                        \end{aligned}
  \end{aligned}
\end{equation}
We can evaluate the traces to be

\begin{subequations}
  \begin{align}
    \Tr\left\{b_kb_{k^\prime}^\dagger\rho_\text{B}\right\} &= \delta_{k k^\prime}\left(1+\bar{n}(\Omega_k)\right)\,, \label{eq:Tr(beta-beta^dagger-rho_B)} \\
    \Tr\left\{b_kb_{k^\prime}\rho_\text{B}\right\} &= 0\,, \\
    \Tr\left\{b_k^\dagger b_{k^\prime}\rho_\text{B}\right\} &= \delta_{k k^\prime}\bar{n}(\Omega_k)\,, \\
    \Tr\left\{b_k^\dagger b_{k^\prime}^\dagger\rho_\text{B}\right\} &= 0\,. \label{eq:Tr(beta^dagger-beta^dagger-rho_B)}
  \end{align}
\end{subequations}
The $\bar{n}(\Omega_k)$ is the Bose-Einstein distribution with temperature of the bath.

With these traces we can evaluate the bath correlation function to become
\begin{multline}
  \label{eq:bath-corr-func-final}
  \left\langle\tilde{B}^\dagger(\tau)\tilde{B}(0)\right\rangle = \sum_{k=0}^\infty |f_{k}|^2\big[ \left(1+\bar{n}(\Omega_k)\right)\e^{-\i\Omega_{k}\tau} \\
    + \bar{n}(\Omega_k)\e^{\i\Omega_{k}\tau} \big]\,.
\end{multline}
Let us now write this in terms of the spectral density as an integral
\begin{multline}
  \label{eq:bath-corr-func-sums-as-integrals}
  \left\langle\tilde{B}^\dagger(\tau)\tilde{B}(0)\right\rangle = \int_0^\infty \d\Omega_kJ(\Omega_k)\big[ \left(1+\bar{n}(\Omega_k)\right)\e^{-\i\Omega_{k}\tau} \\
  + \bar{n}(\Omega_k)\e^{\i\Omega_{k}\tau} \big]\,,
\end{multline}
where $J(\Omega)$ is the spectral density, describing the coupling strength of each bath mode to the system continuously. \new{A general form for the spectral function is \cite{breuer2002theory}}
\new{
\begin{equation}
  \label{eq:spec-dens-def}
  J(\omega) = \sum_k |f_k|^2\delta(\omega-\omega_k)\,,
\end{equation}}
\new{where $f_k$ are coefficients determining the coupling strength of each of the bath modes $\omega_k$ to the system.}

\new{In our code the spectral density function can take any user-specified form. For the example of the qubit-resonator chain, for simplicity we} take the spectral density to be Ohmic with a Drude cutoff:
\begin{equation}
  \label{eq:spectral-dens}
  J(\Omega) = \chi\frac{\Omega}{1+\Omega^2/\omega_\text{c}^2}\,,
\end{equation}
with some constant $\chi$ and a cut-off frequency $\omega_\text{c}$.

The 
coefficient $\gamma(\omega, \omega^\prime)$ from Eq.~(\ref{eq:gamma(omega,omega-prime)}) can then be written as
\begin{equation}
  \label{eq:gamma-omega,omega'-expanded-as-integrals}
  \gamma(\omega, \omega^\prime) = \pi \left[I_\beta(\omega) + I_\beta(\omega^\prime)\right] + \i\left[S_\beta(\omega) - S_\beta(\omega^\prime)\right]\,,
\end{equation}
with the following integrals: 
\begin{multline}
  \label{eq:I_beta(omega)-definition}
  I(\omega) \equiv \int_0^\infty \d\omega_kJ(\omega_k)\big[\left(1+\bar{n}(\omega_k)\right)\delta(\omega-\omega_{k}) \\
    + \bar{n}(\omega_k)\delta(\omega+\omega_{k})\big]\,,
\end{multline}
and
\begin{equation}
  \label{eq:S_beta(omega)-definition}
  S(\omega) \equiv \text{P.V.}\int_0^\infty \d\omega_kJ(\omega_k)\left(\frac{1+\bar{n}(\omega_k)}{\omega-\omega_{k}} + \frac{\bar{n}(\omega_k)}{\omega+\omega_{k}}\right)\,.
\end{equation}

Motivated by \cite{Becker2024}, we compute the integrals analytically \mar{using the} residue theorem when the spectral density is of the form in Eq.~(\ref{eq:spectral-dens}).
The results are given by
\begin{equation}
  \label{eq:I(omega)-result}
  I(\omega) = J(\omega)(1 + \bar{n}(\omega)),
\end{equation}
and 
\begin{equation}
  \begin{aligned}
    S(\omega) = &-\frac{\chi}{2 \pi  \left(1 + \frac{\omega^2}{\omega_\text{c}^2}\right)} \\
                &\times
                  \begin{aligned}[t]
                    \bigg(&\pi\omega_\text{c} - \pi\omega \cot \left(\frac{\omega_\text{c}}{2T}\right) - 2\omega\Re\left[H\left(-\frac{i \omega }{2 \pi  T}\right)\right] \\
                    &+ \omega H\left(-\frac{\omega_\text{c}}{2\pi T}\right) + \omega H\left(\frac{\omega_\text{c}}{2 \pi T}\right)\bigg),
                  \end{aligned}
  \end{aligned}
\end{equation}
where $H(x)$ is the harmonic number of $x$ and $\Re[x]$ is the real part of $x$.

\mar{Similarly,} the coefficients (\ref{eq:Pi(omega,omega-prime)}) in the Lamb-shift Hamiltonian (\ref{eq:Lamb-shift-Hamiltonian}) can be constructed as
\begin{equation}
  \label{eq:pi_beta,beta'(omega,omega')-expanded-as-integrals}
  \pi(\omega, \omega^\prime) = \frac{\pi}{2\i}\left[I(\omega) - I(\omega^\prime)\right] + \frac{1}{2}\left[S(\omega) + S(\omega^\prime)\right]\,.
\end{equation}

\bibliographystyle{unsrtnm}
\bibliography{library}

\end{document}